\documentclass[apj,numberedappendix,twocolappendix,appendixfloats]{emulateapj}
\usepackage{apjfonts}
\usepackage[breaklinks,colorlinks,citecolor=blue,linkcolor=magenta]{hyperref} 

\newcommand{\etal}{et~al.}
\newcommand{\PVdblt}{{\rm P}\kern 0.1em{\sc v}~$\lambda\lambda 1117, 1128$}
\newcommand{\CaIIdblt}{{\rm Ca}\kern 0.1em{\sc ii}~$\lambda\lambda 3934, 3969$}
\newcommand{\AlIIIdblt}{{\rm Al}\kern 0.1em{\sc iv}~$\lambda\lambda 1855, 1863$}
\newcommand{\CIVdblt}{{\rm C}\kern 0.1em{\sc iv}~$\lambda\lambda 1548, 1550$}
\newcommand{\MgIIdblt}{{\rm Mg}\kern 0.1em{\sc ii}~$\lambda\lambda 2796, 2803$}
\newcommand{\NVdblt}{{\rm N}\kern 0.1em{\sc v}~$\lambda\lambda 1238, 1242$}  
\newcommand{\SVIdblt}{{\rm S}\kern 0.1em{\sc vi}~$\lambda\lambda 933, 944$} 
\newcommand{\OVIdblt}{{\rm O}\kern 0.1em{\sc vi}~$\lambda\lambda 1031, 1037$} 
\newcommand{\SiIIdblt}{{\rm Si}\kern 0.1em{\sc ii}~$\lambda\lambda 1190, 1193$} 
\newcommand{\SiIVdblt}{{\rm Si}\kern 0.1em{\sc iv}~$\lambda\lambda 1393, 1402$} 
\newcommand{\PV}{\hbox{{\rm P}\kern 0.1em{\sc v}}}
\newcommand{\AlI}{\hbox{{\rm Al}\kern 0.1em{\sc i}}}
\newcommand{\AlII}{\hbox{{\rm Al}\kern 0.1em{\sc ii}}}
\newcommand{\AlIII}{{\hbox{\rm Al}\kern 0.1em{\sc iii}}}
\newcommand{\CaII}{\hbox{{\rm Ca}\kern 0.1em{\sc ii}}}
\newcommand{\CII}{\hbox{{\rm C}\kern 0.1em{\sc ii}}}
\newcommand{\CIIe}{\hbox{{\rm C$^{\ast}$}\kern 0.1em{\sc ii}}}
\newcommand{\CIII}{\hbox{{\rm C}\kern 0.1em{\sc iii}}}
\newcommand{\CIV}{\hbox{{\rm C}\kern 0.1em{\sc iv}}}
\newcommand{\CV}{\hbox{{\rm C}\kern 0.1em{\sc v}}}
\newcommand{\HI}{\hbox{{\rm H}\kern 0.1em{\sc i}}}
\newcommand{\HII}{\hbox{{\rm H}\kern 0.1em{\sc ii}}}
\newcommand{\Lya}{\hbox{{\rm Ly}\kern 0.1em$\alpha$}}
\newcommand{\Lyb}{\hbox{{\rm Ly}\kern 0.1em$\beta$}}
\newcommand{\Lyg}{\hbox{{\rm Ly}\kern 0.1em$\gamma$}}
\newcommand{\Lyd}{\hbox{{\rm Ly}\kern 0.1em$\delta$}}
\newcommand{\Lye}{\hbox{{\rm Ly}\kern 0.1em$\epsilon$}}
\newcommand{\Lyphi}{\hbox{{\rm Ly}\kern 0.1em$\phi$}}
\newcommand{\Lyfive}{\hbox{{\rm Ly}\kern 0.1em$5$}}
\newcommand{\Lysix}{\hbox{{\rm Ly}\kern 0.1em$6$}}
\newcommand{\Lyseven}{\hbox{{\rm Ly}\kern 0.1em$7$}}
\newcommand{\Lyeight}{\hbox{{\rm Ly}\kern 0.1em$8$}}
\newcommand{\Lynine}{\hbox{{\rm Ly}\kern 0.1em$9$}}
\newcommand{\Lyten}{\hbox{{\rm Ly}\kern 0.1em$10$}}
\newcommand{\Lyeleven}{\hbox{{\rm Ly}\kern 0.1em$11$}}
\newcommand{\HeI}{\hbox{{\rm He}\kern 0.1em{\sc i}}}
\newcommand{\HeII}{\hbox{{\rm He}\kern 0.1em{\sc ii}}}
\newcommand{\FeI}{\hbox{{\rm Fe}\kern 0.1em{\sc i}}}
\newcommand{\FeII}{\hbox{{\rm Fe}\kern 0.1em{\sc ii}}}
\newcommand{\FeIII}{\hbox{{\rm Fe}\kern 0.1em{\sc iii}}}
\newcommand{\MnII}{\hbox{{\rm Mn}\kern 0.1em{\sc ii}}}
\newcommand{\MgI}{\hbox{{\rm Mg}\kern 0.1em{\sc i}}}
\newcommand{\MgIb}{\hbox{{\rm Mg}\kern 0.1em{\sc i}}\kern 0.05em{\rm b}}
\newcommand{\MgII}{\hbox{{\rm Mg}\kern 0.1em{\sc ii}}}
\newcommand{\MgIII}{\hbox{{\rm Mg}\kern 0.1em{\sc iii}}}
\newcommand{\NI}{\hbox{{\rm N}\kern 0.1em{\sc i}}}
\newcommand{\NII}{\hbox{{\rm N}\kern 0.1em{\sc ii}}}
\newcommand{\NIII}{\hbox{{\rm N}\kern 0.1em{\sc iii}}}
\newcommand{\NV}{\hbox{{\rm N}\kern 0.1em{\sc v}}}
\newcommand{\OVI}{\hbox{{\rm O}\kern 0.1em{\sc vi}}}
\newcommand{\OI}{\hbox{{\rm O}\kern 0.1em{\sc i}}}
\newcommand{\OII}{\hbox{[{\rm O}\kern 0.1em{\sc ii}]}}
\newcommand{\OIII}{\hbox{[{\rm O}\kern 0.1em{\sc iii}]}}
\newcommand{\OIV}{\hbox{{\rm O}\kern 0.1em{\sc iv}]}}
\newcommand{\SI}{{\rm S}\kern 0.1em{\sc i}}
\newcommand{\SIV}{{\rm S}\kern 0.1em{\sc iv}}
\newcommand{\SVI}{{\rm S}\kern 0.1em{\sc vi}}
\newcommand{\SiI}{\hbox{{\rm Si}\kern 0.1em{\sc i}}}
\newcommand{\SiII}{\hbox{{\rm Si}\kern 0.1em{\sc ii}}}
\newcommand{\SiIII}{\hbox{{\rm Si}\kern 0.1em{\sc iii}}}
\newcommand{\SiIV}{\hbox{{\rm Si}\kern 0.1em{\sc iv}}}
\newcommand{\SII}{\hbox{{\rm S}\kern 0.1em{\sc ii}}}
\newcommand{\SIII}{\hbox{{\rm S}\kern 0.1em{\sc iii}}}
\newcommand{\NaI}{\hbox{{\rm Na}\kern 0.1em{\sc i}}}
\newcommand{\NaID}{\hbox{{\rm Na}\kern 0.1em{\sc i}}\kern 0.05em{\rm D}}
\newcommand{\TiII}{\hbox{{\rm Ti}\kern 0.1em{\sc ii}}}
\newcommand{\kms}{\hbox{~km~s$^{-1}$}}
\newcommand{\cmsq}{\hbox{cm$^{-2}$}}

\slugcomment{} 
\shorttitle{\sc Galaxy \& {\OVI} absorption kinematics}
\shortauthors{\sc Kacprzak et~al.}

\begin{document}


\title{The Relation Between Galaxy ISM and Circumgalactic {\OVI} Gas
  Kinematics Derived from Observations and $\Lambda$CDM Simulations}


\author{\sc Glenn G. Kacprzak\altaffilmark{1}, Jacob R. Vander
  Vliet\altaffilmark{3,1}, Nikole M. Nielsen\altaffilmark{1}, Sowgat
  Muzahid\altaffilmark{2}, Stephanie K. Pointon\altaffilmark{1},
  Christopher W. Churchill\altaffilmark{3}, Daniel
  Ceverino\altaffilmark{4}, Kenz S. Arraki\altaffilmark{3}, Anatoly
  Klypin\altaffilmark{3}, Jane C. Charlton\altaffilmark{5}, James
  Lewis\altaffilmark{3,1}}
                                                                   
\altaffiltext{1}{Swinburne University of Technology, Victoria 3122, Australia {\tt gkacprzak@swin.edu.au}}
\altaffiltext{2}{Leiden Observatory, Leiden University, P.O. Box 9513, 2300 RA Leiden, The Netherlands}
\altaffiltext{3}{New Mexico State University, Las Cruces, NM 88003, USA}
\altaffiltext{4}{Institut f\"ur Theoretische Astrophysik, Zentrum
  f\"ur Astronomie, Universit\"at Heidelberg, Albert-Ueberle-Str. 2,
  D-69120 Heidelberg, Germany}
\altaffiltext{5}{The Pennsylvania State University, State College, PA 16801, USA}

\begin{abstract}
We present the first galaxy--{\OVI} absorption kinematic study for 20
absorption systems (EW>0.1~{\AA}) associated with isolated galaxies
($0.15\leq z\leq0.55$) that have accurate redshifts and rotation
curves obtained using Keck/ESI. Our sample is split into two azimuthal
angle bins: major axis ($\Phi<25^{\circ}$) and minor axis
($\Phi>33^{\circ}$). {\OVI} absorption along the galaxy major axis is
not correlated with galaxy rotation kinematics, with only 1/10 systems
that could be explained with rotation/accretion models. This is in
contrast to co-rotation commonly observed for {\MgII}
absorption. {\OVI} along the minor axis could be modeled by
accelerating outflows but only for small opening angles, while the
majority of the {\OVI} is decelerating. Along both axes, stacked
{\OVI} profiles reside at the galaxy systemic velocity with the
absorption kinematics spanning the entire dynamical range of their
galaxies. The {\OVI} found in AMR cosmological simulations exists
within filaments and in halos of $\sim$50~kpc surrounding
galaxies. Simulations show that major axis {\OVI} gas inflows along
filaments and decelerates as it approaches the galaxy while increasing
in its level of co-rotation. Minor axis outflows in the simulations
are effective within 50-75~kpc beyond that they decelerate and fall
back onto the galaxy.  Although the simulations show clear {\OVI}
kinematic signatures they are not directly comparable to
observations. When we compare kinematic signatures integrated through
the entire simulated galaxy halo we find that these signatures are
washed out due to full velocity distribution of {\OVI} throughout the
halo. We conclude that {\OVI} alone does not serve as a useful
kinematic indicator of gas accretion, outflows or star-formation and
likely best probes the halo virial temperature.
\end{abstract}




\section{Introduction}
\label{sec:intro}

The circumgalactic medium is a massive reservoir of multi-phased gas
extending out to 200~kpc and reflects the ongoing physical processes
of galaxy evolution.  The CGM makes up as much as 50\% of baryons
around galaxies \citep{tumlinson11,werk14} and the amount of {\OVI}
within the CGM is significant
\citep{stocke06,tumlinson11,fox13,stocke13,peeples14,werk14} with the
vast majority of it bound to the galaxy's gravitational potential
\citep{tumlinson11,stocke13,mathes14}. However, we are yet to
understand the origins and sources of {\OVI} absorption.

It is well known that the {\OVI} equivalent width is anti-correlated
with the projected separation from the host galaxy
\citep[e.g.,][]{tripp08,wakker09,chen09,prochaska11,tumlinson11,
  mathes14,johnson15,kacprzak15}. This is similar to the
anti-correlation observed between {\MgII} equivalent width and impact
parameter
\citep[e.g.,][]{bb91,steidel95,bouche06,kacprzak08,chen10a,bordoloi11,
  nielsen13b,kacprzak13,lan14,lan18,lopez18,rubin18}.  Both {\OVI} and
{\MgII} exhibits a bi-modal azimuthal angle distribution, suggesting a
co-spatial behavior and possibly a kinematic connection or origin
\citep{bouche12,kacprzak12a,kacprzak15}.

It is clear now that the galaxy--{\MgII} absorption relationship shows
strong kinematic preferences consistent with large-scale outflows
\citep{bouche06,tremonti07,martin09,weiner09,chelouche10,
  nestor11,noterdaeme10,coil11,kacprzak10,kacprzak14,rubin10,menard12,
  martin12,noterdaeme12,krogager13,peroux13,rubin14,crighton15,
  nielsen15,nielsen16} and co-rotation/accretion \citep[see][for
review]{kacprzak17}. Kinematically however, we do not know how {\OVI}
relates to its host galaxy.  

It is clear that the kinematics of the {\MgII} and {\OVI} absorption
profiles can be very different in shape and velocity spread or they
can sometimes be similar \citep[e.g.,][]{werk16,nielsen17}.
Examination of the absorption line profile kinematics and column
density ratios has shown that low, intermediate, and high ions may all
have a photoionized origin \citep{tripp08,muzahid15,pachat16}, while
sometimes {\OVI} is commonly found to have a collisionally ionized
origin \citep{tumlinson05,fox09, savage11,tripp11,
  kacprzak12b,narayanan12,wakker12,meiring13,narayanan18,rosenwasser18}.
This implies that {\OVI} can trace warm/hot coronal regions
surrounding galaxies, which may dictate the formation and destruction
of the cool/warm CGM \citep{mo96,maller04,dekel06} or trace other
multi-phase gas structures. Simulations further predict that the
{\OVI} may be directly sensitive to the galaxy halo virial
temperatures, where {\OVI} peaks for L$^{*}$ galaxies
\citep{oppenheimer16} or due to black hole feedback impacting the
physical state of the circumgalactic medium
\citep{nelson18,oppenheimer18}. In addition, \citet{roca-fabrega18}
showed that {\OVI} not only depends on mass but on redshift as
well. Photoionization of cold-warm gas dominates during the peak of
the meta-galactic UV background ($z=2$). In massive halos, collisional
ionization by thermal electrons become important at $z<0.5$.

Thus, although {\MgII} and {\OVI} exhibit some similarities, their
differences make it completely unclear as to whether {\MgII} and
{\OVI} are even trace the same kinematic structures.

We aim to further explore the multi-phase azimuthal distribution of
{\OVI} absorption to determine whether the relative galaxy-{\OVI}
kinematics shows signatures of inflow and outflow along the major and
minor axes, respectively. We have acquired Keck/ESI spectra for 20
galaxies to obtain their rotation curves, which will then be compared
to the {\it HST}/COS {\OVI} absorption kinematics.  In
Section~\ref{sec:data} we present our sample, data and data
reduction. In Section~\ref{sec:results} we present our observational
results and simple models for {\OVI} residing along the major and
minor axes of galaxies.  We provide our interpretation of the data
using cosmological simulations in Section~\ref{sec:sim}.  In
Section~\ref{sec:discussion}, we discuss what can be inferred from the
results and concluding remarks are offered in
Section~\ref{sec:conclusion}.  Throughout we adopt an H$_{\rm
  0}=70$~\kms Mpc$^{-1}$, $\Omega_{\rm M}=0.3$, $\Omega_{\Lambda}=0.7$
cosmology.

\section{GALAXY SAMPLE AND DATA ANALYSIS}
\label{sec:data}

We have obtained rotation curves using Keck/ESI for a sample of 20
{\OVI} absorbing galaxies with redshifts ranging between
0.15$<$$z$$<$0.55 within $\sim300$~kpc (31$<$$D$$<$276~kpc) of bright
background quasars.  These galaxies are selected to be isolated such
that there are no neighbors within 100~kpc and have velocity
separations less than 500~{\kms}. These {\it HST} imaged
galaxy--absorber pairs were identified as part of our ``Multiphase
Galaxy Halos'' Survey (from PID 13398 plus from the literature).  We
discuss the data and analysis below.

\subsection{Quasar Spectroscopy}

The {\it HST}/COS quasar spectra have a resolution of $R\sim$20,000
and covers the {\OVIdblt} doublet for the targeted galaxies. Details
of the {\it HST}/COS observations are presented in \citet{kacprzak15}.
The data were reduced using the {\sc CALCOS} software. Individual
grating integrations were aligned and co-added using the {\sc IDL}
code `coadd\_x1d' developed by
\citet{danforth10}\footnote{http://casa.colorado.edu/danforth/science/cos/costools.html}.
Since the COS FUV spectra are over-sampled (six pixels per resolution
element) we binned the data by three pixels to increase the
signal-to-noise ratio and all of our analysis was performed on the
binned spectra. Continuum normalization was performed by fitting the
absorption-free regions with smooth low-order polynomials.

We adopted the fitted rest-frame equivalent widths (EWs) and column
densities from \citet{kacprzak15}. Non-Gaussian line spread functions
(LSF) were adopted and were obtained by interpolating the LSF tables
\citep{kriss11} at the observed central wavelength for each absorption
line and was convolved with the fitted model Voigt profile {\sc VPFIT}
\citep{carswell14}. In all cases, a minimum number of components was
used to obtain a satisfactory fit with reduced $\chi^2 \sim 1$. The
{\OVI} $\lambda$~1031 model profiles were used to compute the EWs and
the 1$\sigma$ errors were computed using the error spectrum. Both the
EWs and column densities are listed in Table~\ref{tab:morph}.

\subsection{HST Imaging and Galaxy Models}

All quasar/galaxy fields have been imaged with {\it HST} using either
ACS, WFC3 or WFPC2. Details of the observations are found in
\citet{kacprzak15} and the filters used are found in
Table~\ref{tab:morph}. ACS and WFC3 data were reduced using the
DrizzlePac software \citep{gonzaga12}. When enough frames were
present, cosmic rays were removed during the multidrizzle process
otherwise, L.A.Cosmic was used \citep{vandokkum01}. WFPC--2 data were
reduced using the WFPC2 Associations Science Products Pipeline (WASPP)
\citep[see][]{kacprzak11b}.

Galaxy photometry was adopted from \citet{kacprzak15}, who used the
Source Extractor software \citep[SExtractor;][]{bertin96} with a
detection criterion of 1.5~$\sigma$ above background.  The $m_{HST}$
magnitudes in each filter are quoted in the AB system and are listed
in Table~\ref{tab:morph}.

We adopt calculated halo masses and virial radii from \citet{ng18},
who applied halo abundance matching methods in the Bolshoi N-body
cosmological simulation \citep{klypin11}; see
\citet{churchill13a,churchill13b} for further details.

The galaxy morphological parameters and orientations are adopted from
\citet{kacprzak15}. In summary, morphological parameters were
quantified by fitting a two-component disk$+$bulge model using GIM2D
\citep{simard02}, where the disk component has an exponential profile
while the bulge has a S{\'e}rsic profile \citep{sersic68} with
$0.2\leq n\leq 4.0$.  The galaxy properties are listed in
Table~\ref{tab:morph}. We use the standard convention of the azimuthal
angle $\Phi=0^{\circ}$ to be along the galaxy projected major axis and
$\Phi=90^{\circ}$ to be along the galaxy projected minor axis.

\begin{deluxetable*}{llccccrrrrrr}
\tabletypesize{\scriptsize}
\tablecaption{Absorption and host galaxy properties\label{tab:morph}}
\tablecolumns{12}
\tablewidth{0pt} 
\tablehead{
\colhead{Quasar}&
\colhead{$z_{\rm abs}$} &
\colhead{$z_{\rm gal}$\tablenotemark{ a}}&
\colhead{HST} &
\colhead{$m_{HST}$} &
\colhead{log(M$_{h}$)} &
\colhead{R$_{\rm vir}$} &
\colhead{$D$} &
\colhead{$i$} &
\colhead{$\Phi$} &
\colhead{EW$_r$} &
\colhead{log $N$({\OVI})}\\
\colhead{field }&
\colhead{ } &
\colhead{}&
\colhead{Filter} &
\colhead{(AB)} &
\colhead{(M$_{\odot}$)} &
\colhead{(kpc)} &
\colhead{(kpc)} &
\colhead{(degree) } & 
\colhead{(degree) } & 
\colhead{(\AA) } & 
\colhead{ }
}
\startdata
\cutinhead{quasar sight-lines located along the galaxy's major axis ($\Phi<25^{\circ}$)}
J035128.54$-$142908.7 &0.356825  & 0.356992   &  F702W  & 20.7 &$12.0_{-0.2}^{+0.3}$& $191_{-26}^{+48}$   & $72.3\pm0.4$  & $28.5_{-12.5}^{+19.8}$ & $4.9_{-40.2}^{+33.0}$  & $0.396\pm0.013$ & $14.76\pm0.17$   \\[+0.3ex]    
J091440.39$+$282330.6 &0.244098  & 0.244312   &  F814W  & 19.6 &$11.9_{-0.2}^{+0.3}$& $171_{-24}^{+49}$   & $105.9\pm0.1$ & $39.0_{-0.2}^{+0.4}$   & $18.2_{-1.0}^{+1.1}$   & $0.333\pm0.028$ & $14.65\pm0.07$   \\[+0.3ex]    
J094331.61$+$053131.4 &0.353286  & 0.353052   &  F814W  & 21.2 &$11.7_{-0.2}^{+0.4}$& $147_{-22}^{+54}$   & $96.5\pm0.3$  & $44.4_{-1.2}^{+1.1}$   & $8.2_{-5.0}^{+3.0}$    & $0.220\pm0.024$ & $14.66\pm0.07$   \\[+0.3ex]    
J095000.73$+$483129.3 &0.211757  & 0.211866   &  F814W  & 18.0 &$12.4_{-0.2}^{+0.2}$& $247_{-29}^{+36}$   & $93.6\pm0.2$  & $47.7_{-0.1}^{+0.1}$   & $16.6_{-0.1}^{+0.1}$   & $0.211\pm0.019$ & $14.32\pm0.04$   \\[+0.3ex]    
J104116.16$+$061016.9 &0.441630  & 0.442173   &  F702W  & 20.9 &$12.0_{-0.2}^{+0.3}$& $193_{-25}^{+42}$   & $56.2\pm0.3$  & $49.8_{-5.2}^{+7.4}$   & $4.3_{-1.0}^{+0.9}$    & $0.368\pm0.023$ & $14.64\pm0.18$   \\[+0.3ex]    
J113910.79$-$135043.6 &0.204297  & 0.204194  &  F702W  & 20.0 &$11.7_{-0.2}^{+0.4}$& $146_{-22}^{+52}$   & $93.2\pm0.3$  & $81.6_{-0.5}^{+0.4}$   & $5.8_{-0.5}^{+0.4}$    & $0.231\pm0.009$ & $14.40\pm0.28$   \\[+0.3ex]    
J132222.46$+$464546.1 &0.214320  & 0.214431   &  F814W  & 18.6 &$12.1_{-0.2}^{+0.3}$& $205_{-26}^{+44}$   & $38.6\pm0.2$  & $57.9_{-0.2}^{+0.1}$    & $13.9_{-0.2}^{+0.2}$   & $0.354\pm0.024$ & $14.62\pm0.12$   \\[+0.3ex]    
J134251.60$-$005345.3 &0.227196  & 0.227042   &  F814W  & 18.2 &$12.4_{-0.2}^{+0.2}$& $252_{-29}^{+36}$   & $35.3\pm0.2$  & $10.1_{-10.1}^{+0.6}$   & $13.2_{-0.4}^{+0.5}$   & $0.373\pm0.023$ & $14.58\pm0.11$   \\[+0.3ex]    
J213135.26$-$120704.8 &0.430164  & 0.430200   &  F702W  & 20.7 &$12.0_{-0.2}^{+0.3}$& $200_{-25}^{+42}$   & $48.4\pm0.2$  & $48.3_{-3.7}^{+3.5}$    & $14.9_{-4.9}^{+6.0}$   & $0.385\pm0.013$ & $14.60\pm0.05$   \\[+0.3ex]    
J225357.74$+$160853.6 &0.390705  & 0.390013   &  F702W  & 20.6 &$12.2_{-0.2}^{+0.2}$& $217_{-28}^{+45}$   & $276.3\pm0.2$ & $76.1_{-1.2}^{+1.1}$    & $24.2_{-1.2}^{+1.2}$   & $0.173\pm0.030$ & $14.29\pm0.04$    \\[+0.3ex] 
\cutinhead{quasar sight-lines located along the galaxy's minor axis ($\Phi>33^{\circ}$)}
J012528.84$-$000555.9 &0.399090  &0.398525    &  F702W  & 19.7 &$12.5_{-0.2}^{+0.2}$& $285_{-32}^{+37}$   & $163.0\pm0.1$ & $63.2_{-2.6}^{+1.7}$   & $73.4_{-4.7}^{+4.6}$   & $0.817\pm0.023$ & $15.16\pm0.04$   \\[+0.3ex]    
J045608.92$-$215909.4 &0.381514  &0.381511    &  F702W  & 20.7 &$12.0_{-0.2}^{+0.3}$& $192_{-26}^{+48}$   & $103.4\pm0.3$ & $57.1_{-2.4}^{+19.9}$  & $63.8_{-2.7}^{+4.3}$   & $0.219\pm0.013$ & $14.34\pm0.13$   \\[+0.3ex]    
J094331.61$+$053131.4 &0.548769  &0.548494    &  F814W  & 21.0 &$12.0_{-0.2}^{+0.3}$& $191_{-25}^{+43}$   & $150.9\pm0.6$ & $58.8_{-1.1}^{+0.6}$   & $67.2_{-1.0}^{+0.9}$   & $0.275\pm0.050$ & $14.51\pm0.07$   \\[+0.3ex]    
J100902.07$+$071343.9 &0.227851  &0.227855    &  F625W  & 20.1 &$11.8_{-0.2}^{+0.4}$& $155_{-23}^{+51}$   & $64.0\pm0.8$  & $66.3_{-0.9}^{+0.6}$   & $89.6_{-1.3}^{+1.3}$   & $0.576\pm0.021$ & $15.14\pm0.10$   \\[+0.3ex]    
J113910.79$-$135043.6 &0.212237  &0.212259    &  F702W  & 20.0 &$11.7_{-0.2}^{+0.4}$& $150_{-22}^{+52}$   & $174.8\pm0.1$ & $85.0_{-0.6}^{+0.1}$   & $80.4_{-0.5}^{+0.4}$   & $0.137\pm0.009$ & $14.12\pm0.12$   \\[+0.3ex]    
J113910.79$-$135043.6 &0.319167  &0.319255    &  F702W  & 20.6 &$11.9_{-0.2}^{+0.3}$& $170_{-24}^{+51}$   & $73.3\pm0.4$  & $83.4_{-1.1}^{+1.4}$   & $39.1_{-1.7}^{+1.9}$   & $0.255\pm0.012$ & $14.41\pm0.09$   \\[+0.3ex]    
J124154.02$+$572107.3 &0.205538  &0.205267    &  F814W  & 19.9 &$11.6_{-0.2}^{+0.4}$& $140_{-21}^{+52}$   & $21.1\pm0.1$  & $56.4_{-0.5}^{+0.3}$   & $77.6_{-0.4}^{+0.3}$   & $0.519\pm0.018$ & $14.89\pm0.13$   \\[+0.3ex]    
J155504.39$+$362847.9 &0.189033  &0.189201    &  F814W  & 18.5 &$12.1_{-0.2}^{+0.3}$& $194_{-25}^{+45}$   & $33.4\pm0.1$  & $51.8_{-0.7}^{+0.7}$   & $47.0_{-0.8}^{+0.3}$   & $0.385\pm0.033$ & $14.74\pm0.17$   \\[+0.3ex]    
J225357.74$+$160853.6 &0.153821  &0.153718    &  F702W  & 19.3 &$11.6_{-0.2}^{+0.5}$& $130_{-20}^{+53}$   & $31.8\pm0.2$  & $59.6_{-1.7}^{+0.9}$   & $33.3_{-1.9}^{+2.7}$   & $0.263\pm0.056$ & $14.59\pm0.06$    \\[+0.3ex]               %
J225357.74$+$160853.6 &0.352708  &0.352787    &  F702W  & 20.3 &$11.9_{-0.2}^{+0.3}$& $180_{-25}^{+50}$   & $203.2\pm0.5$ & $36.7_{-4.6}^{+6.9}$   & $88.7_{-4.8}^{+4.6}$   & $0.381\pm0.036$ & $14.70\pm0.15$       
\enddata 
 \tablenotetext{a}{Keck ESI redshifts derived from this work.}
\end{deluxetable*} 
 \subsection{Galaxy Spectroscopy}

The galaxy spectra were obtained using the Keck Echelle Spectrograph
and Imager, ESI, \citep{sheinis02}.  The ESI slit position angle was
selected to be near the optical major axis of each galaxy in order to
accurately measure the galaxy rotation curves. Details of the ESI/Keck
observations are presented in Table~\ref{tab:ESI}. The ESI slit is
$20''$ in length and set to $1''$ wide. We binned by two in the
spatial directions resulting in pixel scales of $0.27-0.34''$ over the
echelle orders of interest. Binning by two in the spectral direction
results in a sampling rate of 22~\kms~pixel$^{-1}$ (${\rm
  FWHM}\sim90$~km/s).  ESI has a wavelength coverage of 4000 to
10,000~{\AA}, which covers multiple emission lines such as {\OII}
doublet, $\rm{H}\beta$, {\OIII} doublet, $\rm{H}\alpha$, and [\NII]
doublet.

All ESI data were reduced using IRAF. Galaxy spectra are both vacuum
and heliocentric velocity corrected to provide a direct comparison
with the absorption line spectra.  The derived wavelength solution was
verified against a catalog of known sky-lines which resulted a rms
difference of $\sim0.03$~{\AA} ($\sim2$~{\kms}).

The galaxy rotation curve extraction was performed following the
methods of \citet{kacprzak10,kacprzak11a} \citep[also
  see][]{vogt96,steidel02}. We extracted multiple spectra along the
spatial direction of a galaxy using three-pixel-wide apertures
(corresponding to approximately one resolution element of
$0.81-1.02''$ for ESI) at one pixel spatial increments along the slit.
To obtain accurate wavelength calibrations, we extract spectra of the
arc lines at the same spatial pixels as the extracted galaxy spectra.
Fitted arc lamp exposures provided a dispersion solution with an RMS
of $\sim0.035$~{\AA} ($\sim2$~{\kms}). The Gaussian fitting algorithm
\citep[FITTER: see][]{archiveI} was used to compute best-fit emission-
and absorption-line centers and widths to derive galaxy redshifts and
kinematics.  Galaxy redshifts were computed at the velocity centroid
of the line, accounting for emission-line resolved kinematics and/or
luminosity asymmetries.  The galaxy redshifts are listed in
Table~\ref{tab:morph}; their accuracy ranges from $3-20$~\kms. The 20
rotation curves are presented in Appendix~\ref{sec:A} for the 10
quasar sight-lines along the galaxy's major axis
(Figures~\ref{fig:A1}--\ref{fig:A5}) and in Appendix~\ref{sec:B} for
the 10 quasar sight-lines along the galaxy's minor axis
(Figures~\ref{fig:B1}--\ref{fig:B5}).

\begin{deluxetable}{lccccc}
\tabletypesize{\scriptsize}
\tablecaption{ESI Observations\label{tab:ESI}}
\tablecolumns{6}
\tablewidth{0pt} 
\tablehead{
\colhead{Quasar}&
\colhead{$z^{'}_{\rm gal}$} &
\colhead{$z^{'}_{\rm gal}$}&
\colhead{Observation} &
\colhead{Slit PA} &
\colhead{Exp}\\
\colhead{field }&
\colhead{} &
\colhead{ref\tablenotemark{ a}}&
\colhead{date} &
\colhead{(deg)} &
\colhead{(sec)} 
}
\startdata
J012528.84$-$000555.9 &   0.3985 & 3  &2014$-$12$-$13   &15     &1800    \\[+0.3ex]    
J035128.54$-$142908.7 &   0.3567 & 1  &2014$-$12$-$13   &110    &2550    \\[+0.3ex]    
J045608.92$-$215909.4 &   0.3818 & 1  &2014$-$12$-$13   &110    &1200    \\[+0.3ex]    
J091440.39$+$282330.6 &   0.2443 & 2  &2016$-$01$-$15   &23     &1500    \\[+0.3ex]    
J094331.61$+$053131.4 &   0.3530 & 2  &2016$-$01$-$15   &38     &4500    \\[+0.3ex]    
J094331.61$+$053131.4 &   0.5480 & 2  &2016$-$01$-$15   &$-$218 &4500    \\[+0.3ex]    
J095000.73$+$483129.3 &   0.2119 & 2  &2016$-$01$-$15   &13     &1000    \\[+0.3ex]    
J100902.07$+$071343.9 &   0.2278 & 2  &2016$-$01$-$15   &115    &1200    \\[+0.3ex]    
J104116.16$+$061016.9 &   0.4432 & 5  &2014$-$04$-$25   &$-$110 &3300    \\[+0.3ex]    
J113910.79$-$135043.6 &   0.2044 & 1  &2016$-$01$-$15   &160    &1650    \\[+0.3ex]    
J113910.79$-$135043.6 &   0.2123 & 1  &2016$-$01$-$15   &111    &2400    \\[+0.3ex]    
J113910.79$-$135043.6 &   0.3191 & 1  &2016$-$06$-$06   &94     &1800    \\[+0.3ex]    
J124154.02$+$572107.3 &   0.2053 & 2  &2016$-$06$-$06   &121    &1200    \\[+0.3ex]    
J132222.46$+$464546.1 &   0.2142 & 2  &2016$-$06$-$06   &0      &1500    \\[+0.3ex]    
J134251.60$-$005345.3 &   0.2270 & 2  &2016$-$06$-$06   &$-$24  &1800    \\[+0.3ex]    
J155504.39$+$362847.9 &   0.1893 & 2  &2016$-$06$-$06   &130    &1800    \\[+0.3ex]    
J213135.26$-$120704.8 &   0.4300 & 6  &2015$-$07$-$16   &55     &6000    \\[+0.3ex]    
J225357.74$+$160853.6 &   0.1530 & 1  &2015$-$07$-$16   &$-$213 &3000    \\[+0.3ex]               %
J225357.74$+$160853.6 &   0.3526 & 1  &2016$-$06$-$06   &$-$185 &3300    \\[+0.3ex]  
J225357.74$+$160853.6 &   0.3900 & 1  &2016$-$06$-$06   &30     &1200  
\enddata 
  \tablenotetext{a}{ Original galaxy redshift ($z^{'}_{\rm gal}$) source: 1) \citet{chen01},  2) \citet{werk12} , 3) \citet{muzahid15}, 4) \citet{kacprzak10}, 5) \citet{steidel02}, 6) \citet{gb97}.}
\end{deluxetable} 

\begin{figure}
\begin{center}
\includegraphics[angle=0,scale=0.43]{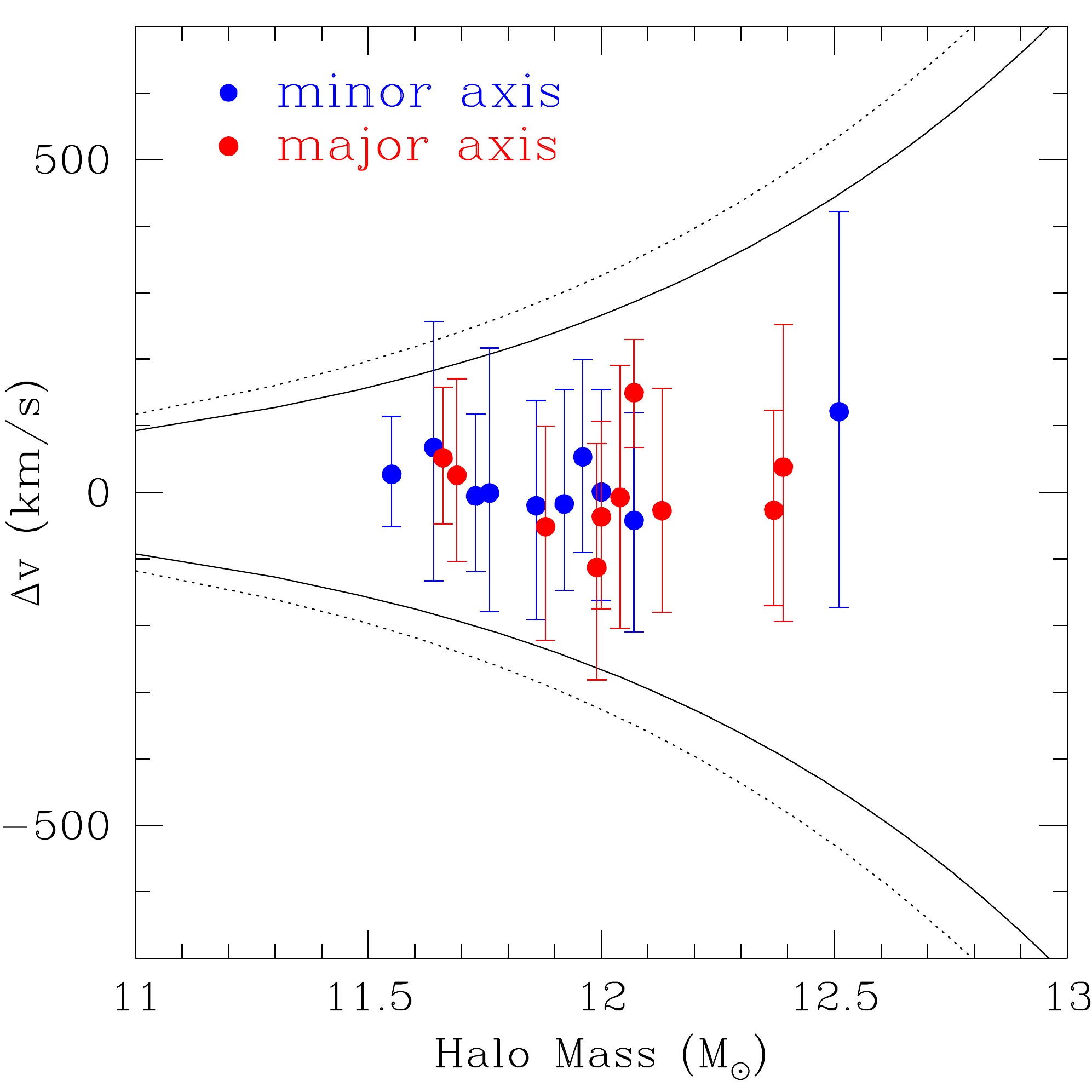}
\caption[angle=0]{The points show the velocity of the median optical
  depth of {\OVI} absorption with respect to their associated galaxy
  systemic velocity as a function of the inferred galaxy halo
  mass. The error-bars indicate the full velocity width of the
  absorption profiles.  Our sample is split into two azimuthal angle
  bins: major axis (red -- $\Phi<25^{\circ}$) and minor axis (blue --
  $\Phi>33^{\circ}$). The two curves indicate the halo escape
  velocities from 100~kpc (dotted) and 200~kpc (solid). Note that
  almost all {\OVI} absorption has velocities well within the halo
  escape velocities.}
\label{fig:escape}
\end{center}
\end{figure}

 \begin{figure*}
\begin{center}
\includegraphics[scale=0.47]{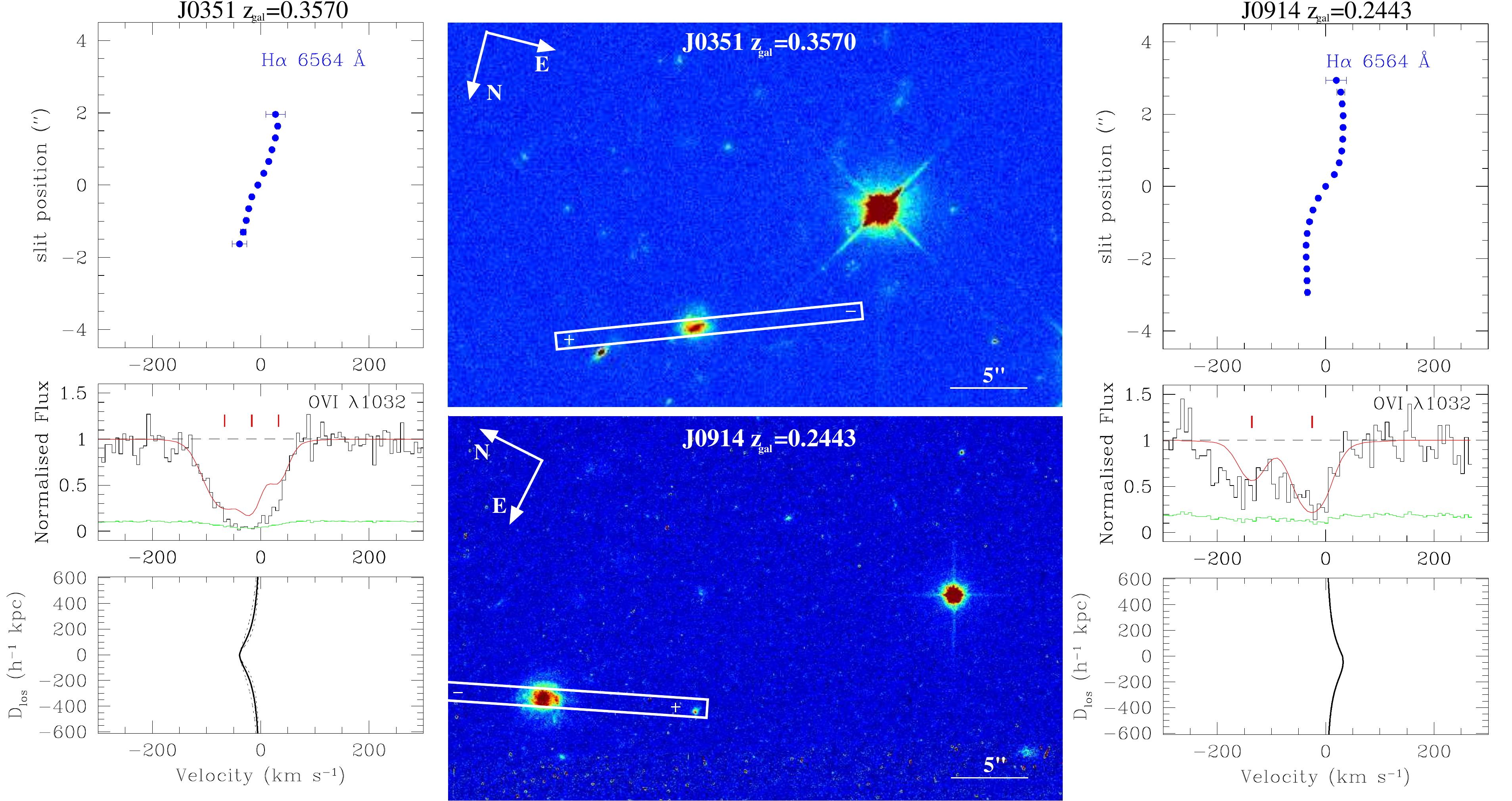}
\caption[angle=0]{{\it HST} images and galaxy rotation curves
  presented for two fields where the quasar sight-line aligns with the
  galaxy's major axis. (Top middle) A 45$'' \times $25$''$ {\it HST}
  image of the quasar field J0351. The ESI/Keck slit is superimposed
  on the image over the targeted galaxy. The "$+$" and "$-$" on the
  slit indicate slit direction in positive and negative arcseconds
  where 0$''$ is defined at the galaxy center.  (Left) The $z=0.3570$
  galaxy rotation curve and the {\it HST}/COS {\OVI} $\lambda$1031
  absorption profile is shown with respect to the galaxy systemic
  velocity. The panel below the {\OVI} absorption is a simple disk
  rotation model computed using Equation~\ref{eq:kine}, which is a
  function of the galaxy rotation speed and orientation with respect
  to the quasar sight-line. The J0351 galaxy is rotating in the same
  direction as the absorption however, the velocity range covered by
  the model is not consistent with the entire range covered by the
  absorption profile.  (Bottom middle) Same as top middle except for
  the J0914 quasar field and for the targeted galaxy at $z = 0.2443$
  (Right) Same as left except the $z = 0.2443$ in the J0914 quasar
  field. Note here that the {\OVI} absorption is consistent with being
  counter-rotating with respect to the galaxy and again, the model has
  insufficient velocities to account for all the absorption
  kinematics. In both cases disk-rotation does not reproduce the
  observed absorption velocities.  Figures for all galaxies are found
  in Appendix~\ref{sec:A} (major-axis) and Appendix~\ref{sec:B}
  (minor-axis).}
\label{fig:moskine}
\end{center}
\end{figure*}

\section{Results}\label{sec:results}

In this section, we explore the kinematic relationship between {\OVI}
absorption and their host galaxies.

\subsection{Gravitationally Bound OVI}

We first explore whether the {\OVI} CGM gas is gravitationally bound
to their host galaxy dark matter halos. In Figure~\ref{fig:escape}, we
show the velocity difference between the median optical depth
distribution of the {\OVI}$\lambda$1031 absorption line and the galaxy
systemic velocity as a function of the host galaxy halo mass for all
azimuthal angles. The error bars show the full velocity range of the
absorption, which is defined as where the Voigt profile fitted
absorption models return to 1\% from the continuum level. The Voigt
profile models are preferred to define the velocity ranges since some
{\OVI} absorption systems are blended with other ions in the spectra
\citep[see][]{nielsen17} and the data tend to be quite noisy.

The rest-frame velocity differences between galaxies and their
associated {\OVI} absorption has a mean offset of
$dv=9.2\pm58.8$~{\kms} with standard error of the mean of
13.5~{\kms}. This implies that most of the gas resides near the galaxy
systemic velocity regardless of its orientation with respect to the
host galaxy.  Also included in the figure are curves indicating the
escape velocity for a given halo mass at an impact parameter of
$D=200$~kpc (inner curve) and 100~kpc (outer curve) at the median
redshift of $z=0.3$. Note that little-to-no absorption resides outside
of these curves, indicating that the {\OVI} gas is bound to their dark
matter halos. These results are consistent with previous findings
showing bound {\OVI} gas
\citep[e.g.,][]{tumlinson11,stocke13,mathes14}.

\subsection{OVI gas kinematics along the galaxy projected major-axis}
\label{sec:majorrot}

Given the observed {\OVI} azimuthal angle bimodality
\citep{kacprzak15}, our sample can be easily split into two azimuthal
angle bins considered as major and minor axis samples. Here we discuss
a subset of 10 systems where the {\OVI} absorption is detected within
25~degrees of the galaxy major axis. This major axis azimuthal cut was
selected to mimic the {\MgII} major axis sample of \citet{ho17}.

We aim to determine if major-axis {\OVI} displays similar co-rotation
kinematic signatures as commonly seen for {\MgII} absorption
\citep[e.g.,][]{steidel02,kacprzak10,ho17}. In
Figure~\ref{fig:moskine}, we present the data used in this analysis
for two example fields J0351 and J0914. The Figures for the targeted
galaxies where the quasar sightline aligns with their major axis are
located in Appendix~\ref{sec:A}. Figure~\ref{fig:moskine} shows the
{\OVI} host galaxies and the quasars in the {\it HST} images along
with the ESI slit position placed over each galaxy. The figure further
shows the H$\alpha$-derived galaxy rotation curve, obtained from the
ESI spectra, and the {\it HST}/COS {\OVI} absorption profile. All
velocities are shown with respect to the galaxy systemic
velocity. Note that the rotation speeds are low ($\sim$50~{\kms}),
which is expected for these moderately inclined spiral galaxies. For
the galaxy in J0351, the {\OVI} absorption profile covers the entire
kinematic range of the galaxy rotation curve. As for the galaxy in
J0914, the {\OVI} absorption resides mostly to one side of the galaxy
systemic velocity as previously seen for {\MgII} systems. We now
explore if co-rotating/lagging disk models can explain the observed
CGM kinematics.

Similar to previous works, we apply the simple monolithic halo model
from \citet{steidel02} to determine whether an extended disk-like
rotating gas disk (as commonly seen for {\MgII}) is able to reproduce
the observed {\OVI} absorption velocity spread given the galaxy's
rotation speed and relative orientation with respect to the quasar
sightline. In summary, model line-of-sight velocities ($v_{los}$) are
a function of the measurable quantities of impact parameter ($D$),
galaxy inclination angle ($i$), galaxy--quasar position angle ($\Phi$)
and the maximum projected galaxy rotation velocity ($v_{max}$) such
that

\begin{eqnarray}
\label{eq:kine}
 v_{los}&=&\frac{-v_{max}}{\sqrt{1+\left(\displaystyle \frac{y}{p}\right)^2}}\mbox{~}
\exp \left\{-\frac{\left|y-y_{\circ}\right|}{h_{v}\tan i}\right\}\mbox{~~}\mbox{where}, \\
\nonumber\\[2.0ex]
y_{\circ}&=&\frac{D\sin \Phi}{\cos i} \mbox{~~~~~}\mbox{and}\mbox{~~~~~}p=D\cos \Phi\mbox{~},\nonumber
\end{eqnarray}                                                 
where $h_v$ is a free parameter representing the scale height for the
velocity lag of the CGM. Here, we assume a thick disk
($h_v=1000$~kpc), which represents the maximum disk/CGM rotation
scenario. Assuming a maximum disk rotation model is reasonable given
that we do not know how/if the velocity changes with impact parameter
and there is little-to-no velocity gradient along the co-rotating
gaseous structures within the simulations
\citep[e.g.,][]{stewart11,stewart13,stewart17}.

The parameter $y$ is the projected line of sight position above the
disk-plane and $y_o$ is the position at the projected disk
mid-plane. The distance along the sightline relative to the point
where it intersects the projection of the disk mid-plane is
$D_{los}=(y-y_o)/$sin$i$. Thus, $D_{los}=0$~kpc is where the model
line-of-sight intersects the projected mid-plane of the galaxy. Please
see figure 6 from \citet{steidel02} for a visual representation of the
model.

Shown in the bottom panels of Figure~\ref{fig:moskine} are the
line-of-sight velocities through the halo derived for the geometry of
both galaxy--quasar pairs for CGM gas rotating at a maximum velocity
set by the rotation curves (solid curves). The dashed curves indicate
model velocities derived from uncertainties in $i$ and $\Phi$. In most
cases, error values are small such that the dashed curves lie near/on
the solid curves. The $z=0.3570$ galaxy in the J0351 field has most of
the {\OVI} absorption blueward of the galaxy systemic velocity, which
agrees with the direction of rotation of the galaxy and thus model
halo. However, given the galaxy's moderate inclination, the model is
unable to account for the large velocity spread measured for the
absorption profile. Similarly, the $z=0.2443$ galaxy in the J0914
field also has the majority of the {\OVI} blueward of the galaxy
systemic velocity. In this case, the galaxy is consistent with being
counter-rotating with respect to the {\OVI}, with this the model
again failing to reproduce the observed velocity spread.

 \begin{figure*}
\begin{center}
\includegraphics[scale=0.95]{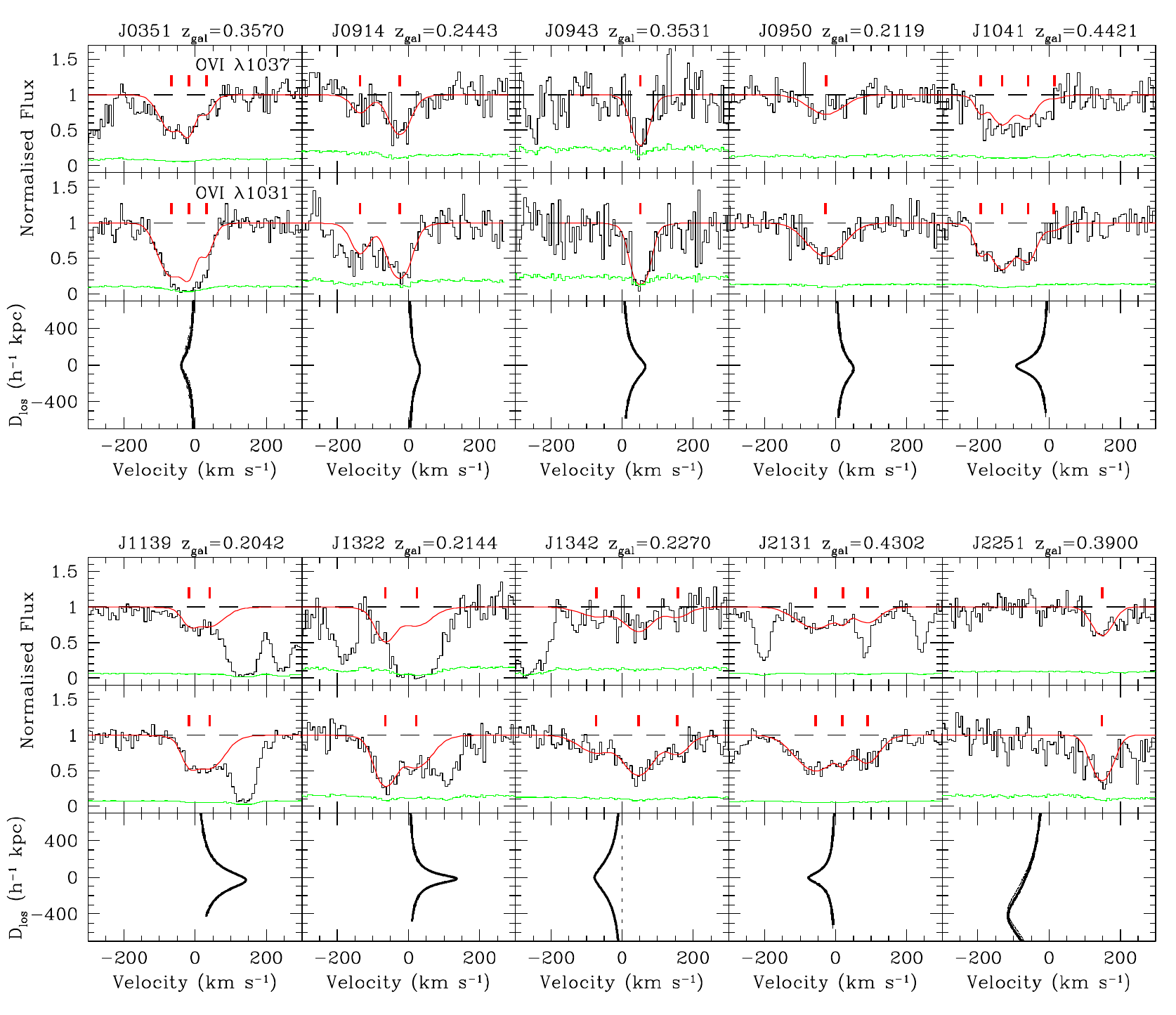}
\caption[angle=0]{{\OVIdblt}~doublet absorption profiles are shown for
  systems where the quasar sight-line is within 25~degrees of the
  galaxy major axis. The red line is a fit to the data and the
  vertical ticks indicate the number of components in each fit.  Also
  shown is the disk model velocities as a function of the distance
  along the sight-line ($D_{los}$). $D_{los}$ is equal to zero when
  the quasar sightline intersects the projected mid-plane of the
  galaxy. The solid curves are computed using Equations~\ref{eq:kine}
  from the values in Table~\ref{tab:morph}. The dashed curves are
  models computed for the maximum and minimum predicted model
  velocities given the uncertainties in $i$ and $\Phi$. The disk model
  is considered successful and reproduces the observations when the
  solid curve overlaps with the bulk of the absorption kinematics.}
\label{fig:mosmajor}
\end{center}
\end{figure*}

 \begin{figure*}
\begin{center}
\includegraphics[scale=0.57]{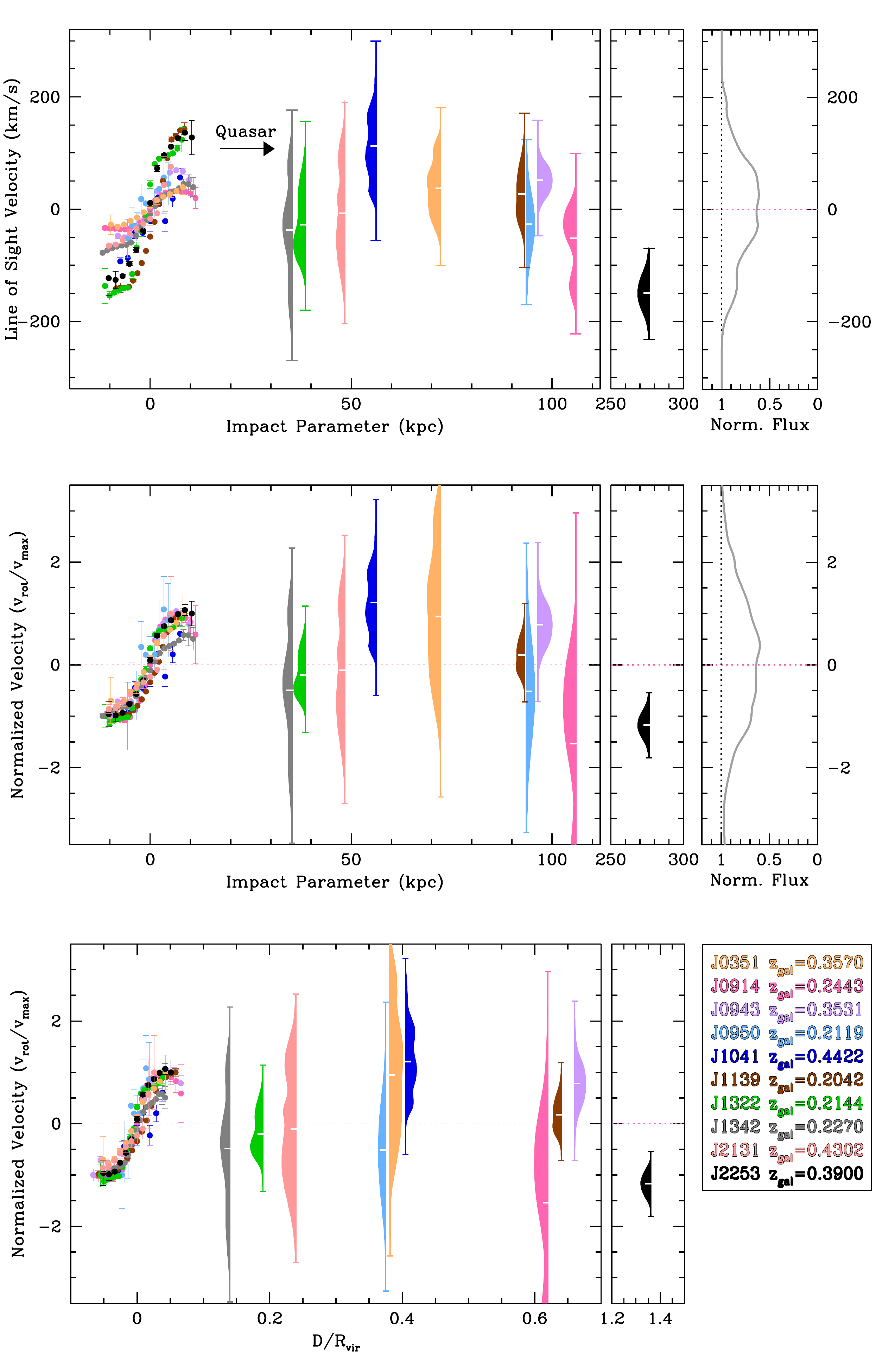}
\caption[angle=0]{ (Top) Rotation curves for galaxies where the quasar
  sightline probes within 25~degrees of the galaxy projected major
  axis as a function of impact parameter. The rotation curves are
  orientated such that quasar is probing gas along the positive
  velocity side of the galaxy. Each galaxy is colored according to the
  key, which is matched with its corresponding absorption profile. The
  {\OVI} is shown with respect to the galaxy systemic velocity and the
  error bars indicate the full extent of the absorption while the
  shaded region shows the optical depth distribution of the actual
  absorption profile in velocity space. The white tickmark on the
  profile indicates the {\OVI} absorption optical depth weighted
  median velocity. Also shown on the right is the average spectrum of
  the 10 absorption profiles. The {\OVI} absorption spans the entire
  velocity range of the galaxy while being centered close to the
  systemic velocity. (Middle) Same as top except that the galaxy
  rotation and absorption velocities are now normalized to the peak
  velocity of each rotation curve. Note that the absorption spans
  twice the dynamical range of the galaxies. (Bottom) Same as middle
  except as a function of the ratio of the impact parameter and the
  inferred halo viral radius.}
\label{fig:vel}
\end{center}
\end{figure*}

Figure~\ref{fig:mosmajor} shows the {\OVIdblt} absorption profiles
along with the Voigt profile fits for the 10 galaxies that have quasar
sightlines passing within 25~degrees of the host galaxy's projected
major axis. The absorption profiles are plotted relative to the host
galaxy systemic velocities. Note that the bulk of the gas resides
near the galaxy systemic velocity with a relatively large velocity
spread. Below the profiles are the modeled co-rotating line-of-sight
velocities. It is immediately clear that the absorption profiles have a
much higher velocity range than that of the predicted model
line-of-sight velocities. Furthermore, only four systems (J0351, J0943,
J1041 and J2131) have models that are rotating in the direction of the
bulk of the {\OVI} but still fall short of predicting the observed
velocities. The $z=0.35$ system in the J0943 field is the only system
where the observed {\OVI} gas could be explained by disk rotation
and/or accretion. Five systems are consistent with counter-rotating
{\OVI} absorption relative to their host galaxies (J0914, J0950,
J1322, J1342 and J2251). These results are in stark contrast from what
has been found for {\MgII} galaxy-absorption pairs.

Given that the quasar sight-lines are within 25~degrees of the galaxy
major axis, we created the top panel of Figure~\ref{fig:vel} shows the
rotation velocities as a function of the projected distance between
the galaxies and their quasar sightlines. The rotation curves for each
galaxy are orientated such that the quasar sightlines are located
along the positive velocity arm of the rotation curves. The {\OVI} is
shown with respect to the galaxy systemic velocity and has the same
color as plotted for the rotation curve of their host galaxy. The
error bars indicate the full extent of the absorption while the shaded
region shows the actual absorption profile in velocity space. This
allows the reader to see where the bulk of the optical depth is in
relation to the galaxy rotation. The white tickmark indicates the
optical depth weighted median of the {\OVI} absorption profile.

It can clearly be seen in the top panel of Figure~\ref{fig:vel} that
the majority of {\OVI} is inconsistent with co-rotation and gas
accretion models.  Most of the gas resides near the galaxy systemic
velocity and there is no preference towards the direction of galaxy
rotation. It is still plausible that some of the gas could be
accreting/co-rotating, however the signature is not strong or is
masked by the other kinematics ongoing within the halo. A mean stacked
spectrum of all 10 absorption systems is also shown in the top panel,
where the {\OVI} absorption almost symmetrically spans the galaxy
systemic velocity where the optical depth weighted median of the
stacked spectrum is at 2.5~{\kms} relative to the galaxy systemic
velocity. Furthermore, the {\OVI} profile encompasses the entire
rotational dynamics of the galaxies.  This is more clearly shown in
the middle panel of Figure~\ref{fig:vel}, where all the velocities are
normalized to the maximum line-of-sight rotation velocity for each
rotation curve and their associated absorption. Almost all of the
systems span the entire dynamical range of the galaxy rotation and
more. Again, a mean stacked spectrum is also shown and the {\OVI} spans
more than twice the full dynamic range of the rotation curves. Recall
though the {\OVI} gas is still gravationally bound to their halos.
The bottom panel in Figure~\ref{fig:vel} is similar to the middle
panel except now shown as a function of viral radius derived for each
galaxy. This clearly shows that 9/10 systems are well within the viral
radius.

These results indicate that a rotating disk and/or radial infall does
not provide a plausible explanation for the total observed {\OVI}
kinematics. Thus, this clearly indicates that if there exits a
kinematic connection between highly ionized gas and its galaxies,
then it is either very low and/or masked by other kinematic sources
such as diffuse gas found within the halo. Given that the quasar
sight-lines are within $<25$ degrees of the galaxy major axes, ongoing
outflows would not likely contribute to the absorption kinematics seen
here. However, it is possible that recycled gas could could be
dominating the observed kinematics.

\subsection{OVI gas kinematics along the galaxy projected minor-axis}
\label{sec:minor}
 
Here we discuss a subset of 10 systems where the {\OVI} absorption is
detected at $>$33~degrees from the galaxy major axis (within 57
degrees of the galaxy minor axis). This angle was selected given that
{\OVI} outflowing gas could likely occur within half-opening angles as
small as 30 degrees or even larger to 50 degrees \citep{kacprzak15}.
Figures~\ref{fig:B1}--\ref{fig:B5} show the {\it HST} images along
with the galaxy rotation curves and their corresponding {\OVI}
absorption. Inspection of these figures shows that the absorption
spans the entire galaxy systemic velocity and encompasses the full
galaxy rotation velocity range in 4/10 cases while 6/10 systems have
most of the {\OVI} absorption offset to one side of the galaxy
systemic velocity.

\begin{figure*}
\begin{center}
\includegraphics[angle=0,scale=0.67]{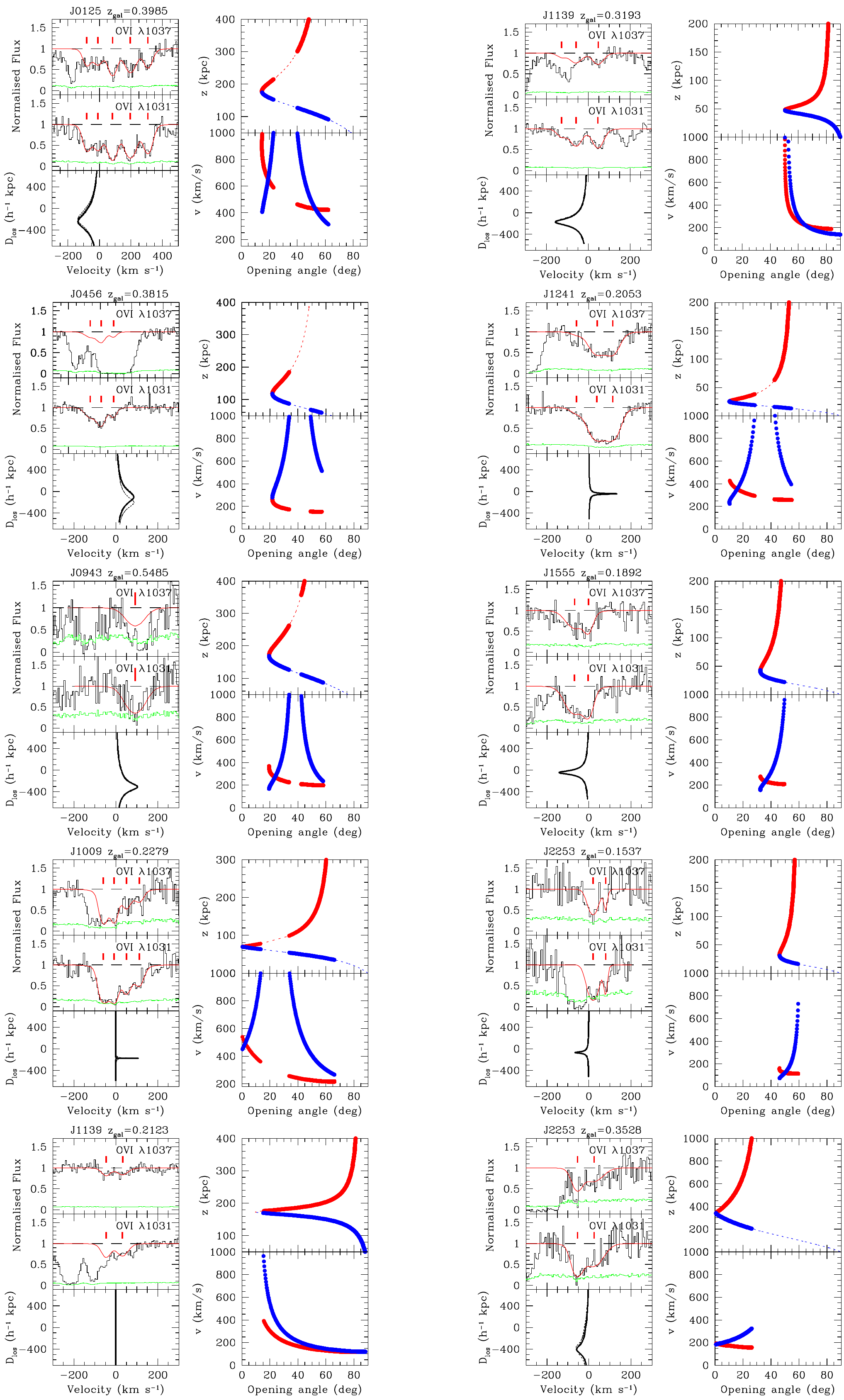}
\caption[angle=0]{(Left) {\OVIdblt} doublet absorption profiles shown
  for systems where the quasar sight-line is within 57~degrees of the
  galaxy minor axis ($\Phi>33$ degrees). The red line is the Voigt
  profile best fit and the vertical ticks indicate the number of
  components.  The panels below the profiles are the disk model
  velocities as a function of the distance along the sight-line
  ($D_{los}$) computed using Equations~\ref{eq:kine} from the values
  in Table~\ref{tab:morph}. The solid and dashed curves are computed
  for the maximum and minimum predicted model velocities given the
  uncertainties in $i$ and $\Phi$. The model is considered successful
  when the solid curve overlaps with the bulk of the absorption
  kinematics. (Right) Outflow models from
  Equations~\ref{eq:flows1}--\ref{eq:flows2} showing the allowed
  parameter space of the $z$-height (top) and outflow velocities
  (bottom) versus the half opening angle. The colored lines highlight
  the height at which the sight lines enter (blue) and leave (red) the
  outflow.  If we assume that the line-of-sight velocity increases
  smoothly from $z_1$ to $z_2$, then outflows accelerate, as seen in
  the lower velocity panel, when the blue line is below the red line
  and decelerate when the red line is below the blue line.}
\label{fig:minorout}
\end{center}
\end{figure*}

Previous studies have shown that have {\MgII} galaxy-absorber pairs
that disk-like rotation can be found for quasar sightlines with
intermediate-to-high $\Phi$ values \citep[e.g.,][]{kacprzak10}. We
explore the disk-like rotation model from Equations~\ref{eq:kine} for
our minor axis sample in Figure~\ref{fig:minorout}. In
Figure~\ref{fig:minorout}, the fitted {\OVI} doublet is shown along
with the modeled line-of-sight velocities through the halo derived for
each galaxy using the maximum velocity set by the rotation curves
(solid curves). We find three systems (J1241, J0943 and J1555) where
the model can account for all the observed absorption
velocities. However, 7/10 have model kinematics consistent with
counter-rotation with respect to the bulk of the {\OVI}
absorption. What these models demonstrate is that there is not an
overall consistency between the relative {\OVI}-galaxy
kinematics. Only 4/20 from the total sample of major and minor axis
galaxies have relative velocities expected of disk-like rotation/gas
accretion.  This is in stark contrast to the commonly observed
co-rotation found for {\MgII} absorption.

Given the relative quasar-galaxy geometry, it could be expected that
outflows might be commonly observed along the galaxy minor
axis. Furthermore, this outflowing gas is likely traced by warm {\OVI}
absorption. To test this, we apply a simple conical model for
outflowing gas from \citet{gauthier12}. Here we summarize the model
but see \citet{gauthier12} for details and their figure~1 for an
illustration of the model.

Their collimated outflow model is characterized by an expanding cone
originating from the galaxy center along the polar axis with a total
angular span of $2\,\theta_0$. As with the disk-rotation model, $i$ is
the inclination of the galaxy while $\Phi$ is the angle between the
projected major axis of the disk and the quasar sightline that is at
an impact parameter $D$. These measured quantities are found in
Table~\ref{tab:morph}. The quasar line-of-sight intercepts the
outer-edges of the outflow cone at a height $z$ from $z_1$ to $z_2$,
which is determined by the cone opening angle $\theta_0$.  The
position angles, $\phi_{[1,2]}$, of the projected outflow
cross-section at $z_{[1,2]}$ are constrained by
\begin{equation}
\label{eq:flows1}
\tan\phi_{[1,2]}=\frac{D\,\sin\alpha-z_{[1,2]}\,\sin i}{D\,\cos\alpha}
\end{equation}
and the relation between  $z_{[1,2]}$ and the opening angle $\theta_0$
is
\begin{equation}
\label{eq:flows2}
z_{[1,2]}\,\tan\theta_0=D\sqrt{1+\sin^2\phi_{[1,2]}\tan^2i}\,\left(\frac{\cos\Phi}{\cos\phi_{[1,2]}}\right).
\end{equation}
Equations~\ref{eq:flows1}--\ref{eq:flows2} can be used to calculate
the corresponding $\theta$ at any given point along the quasar
sightline at height $z$ where $z_1\le z\le z_2$. 

The ouflow speed, $v$, of a gas cloud moving outward at a hight $z$
corresponds to the line-of-sight velocity $v_{\rm los}$ such that
\begin{equation}
v = \frac{v_{\,\rm los}}{\cos j}\mbox{,~~}\mbox{where}\mbox{~~} \\
\nonumber\\[2.0ex]
j = \sin^{-1}\left(\frac{D}{z}\cos\theta\right).
\nonumber
\end{equation}
The line-of-sight velocities are defined by the red-most and blue-most
velocity edges of the absorption profile relative to the galaxy
systemic velocity.  The gas producing the observed absorption is
assumed to be distributed symmetrically around the polar axis of the
cone and the absorption at $z_1$ and $z_2$ probes regions close to the
front and back side of the outflow respectively. If asymmetry arises
due to inhomogeneities of gas with the outflows, then the computed
velocity gradients represent a lower limit to outflow velocity field.

We apply the above model since our observational data of the
galaxy-quasar geometry provides constraints on $\theta_0$ and the
absorption profiles constrain the plausible outflow velocities. From
these data, we can identify whether or not reasonable opening angles
and outflow velocities are able to replicate the observations. If so,
then outflows are a plausible explanation for the observed kinematics,
and if not, then outflows may not be the likely source driving the
{\OVI} gas kinematics seen along the galaxy minor axis.

Figure~\ref{fig:minorout} shows the outflow models for each
quasar-galaxy pair for their relative orientation and absorption
gas-galaxy kinematics. The right top panel for each system shows at
what height the quasar sightline enters the outflow cone (blue -- dash
and full lines) and what height the sight-line exits the cone (red --
dash and full lines) as a function of outflow opening angle.  The
Figure shows scenarios where the opening angle is not well
constrained, as seen for J1139\_0.2123, since the galaxy is nearly
edge-on ($i=85$~degrees) and the quasar sightline almost directly
along the minor axis (within 9.6 degrees).  The other scenario shown
is for galaxies where the quasar sight-line is not directly along the
galaxy minor axis and the outflow opening angle has to be sufficiently
large enough before it intercepts the sight-line. This can be seen for
J1139\_0.3193 ($\Phi=39$ degrees) and for J2253\_0.1537 ($\Phi=33$
degrees) where the opening has to be at least 50 degrees before the
sight-line intercepts the cone.

In all cases, the opening angle on both sides of the cone can be large
enough such that the quasar sightline no longer intercepts the cone,
which is why $z$ asymptotes to large values. We use the far side of
the cone (red) as an upper limit on the outflow half opening
angle. From geometric arguments only, the model constrains the half
opening angles to range from 0--50~degrees as the smallest possible
angle to 26--83 degrees at its largest.  The half opening angle model
results are presented in Table~\ref{tab:model}. These are consistent
with expected/modeled values found for cooler gas tracers, which range
between 10--70 degrees
\citep{walter02,gauthier12,kacprzak12a,martin12,bordoloi14}. These are
also consistent with those derived by \citet{kacprzak15} who examined
the azimuthal angle dependence of the gas covering fraction and
concluded that the {\OVI} outflowing gas could occur within a
half-opening angle as small as 30$^{\circ}$ or even larger at
50$^{\circ}$.

If realistic outflow velocities can reproduce the observed absorption,
it would be a key step for understanding whether outflows can explain
the observed OVI gas-galaxy kinematics. The bottom panels in
Figure~\ref{fig:minorout} show the model outflow velocities at the
edges of the cones required to reproduce the entire velocity spread of
the observed {\OVI} absorption profile with respect to the galaxy
systemic velocity. The blue line corresponds to the outflow velocities
where the quasar sightline enters the outflow cone, while the red line
corresponds to the outflow velocities where the quasar sightline exits
the outflow cone. Each galaxy-absorber pair has a large range of
modeled velocities as a function of opening angle required to
reproduce the observed line-of-sight velocities. Some of these
velocities far exceed 1000~{\kms} as the dot product of the outflow
velocity vector and the line-of-sight velocity vector approaches
zero. Here we assume that the outflow velocities at large distances
above the galaxy disk likely do not exceed 1000~{\kms} for these
systems. This assumption provides additional constraints on the
acceptable outflow geometry indicated by the solid line and those
values are listed in Table~\ref{tab:model}. While the range in opening
angles is more limited, the viable half-opening angles are still
consistent with previous works.

The ranges of the outflows velocities required to reproduce the
observed {\OVI} kinematics shown on Figure~\ref{fig:minorout} appear
reasonable and are within a few hundred {\kms} -- typical of expected
outflow velocities
\citep[e.g.,][]{martin09,weiner09,bouche12,gauthier12,martin12,rubin14,schroetter16}.
We emphasize, however, that a modeled active accelerating outflow
occurs when the line-of-sight velocity increases from where the quasar
sightline enters the conical outflow to where the quasar sightline
exits the conical outflow (since $z_1<z_2$). The outflow velocities
$(v)$ shown in Figure~\ref{fig:minorout} would be consistent with an
active outflowing model when the blue line is below the red line.  In
the opposite case, where the red line is below the blue line, the
outflow is decelerating in order to reproduce the observed
kinematics. 

We find that 7/10 galaxies exhibit outflowing gas with an accelerated
flow. Note however, that acceleration only occurs for very small
opening angles typically within 20~degrees and only over a range of 10
degrees (with the exception of J1139\_0.3193).  These values are also
listed in Table~\ref{tab:model}. Thus, if active outflows are
occurring, they only occur within a very small opening angle conical
outflow.
\begin{deluxetable*}{lccccccc}[!ht]
\tabletypesize{\scriptsize}
\tablecaption{Model outflow half opening angles and velocities\label{tab:model}}
\tablecolumns{8}
\tablewidth{0pt} 
\tablehead{
\colhead{Quasar}&
\colhead{$z_{\rm abs}$} &
\colhead{$z(i)$\tablenotemark{ a}}&
\colhead{$v_{red}(i)$\tablenotemark{ b}} &
\colhead{$v_{blue}(i)$\tablenotemark{ c}} &
\colhead{$\theta$ (deg.)\tablenotemark{ d}} &
\colhead{$\theta$ (deg.)\tablenotemark{ e}} &
\colhead{$\theta$ (deg.)\tablenotemark{ f}}\\
\colhead{field}&
\colhead{ } &
\colhead{(kpc)}&
\colhead{(\kms)} &
\colhead{(\kms)} &
\colhead{(geometric)} &
\colhead{(velocity limited)} &
\colhead{(acceleration only)} 
}
\startdata 
J012528.84$-$000555.9 &0.398525  &   174 &      991  &     406  &  15--62 & 15--23 \& 40.1--62 &15--19 \\[+0.3ex] 
J045608.92$-$215909.4 &0.381511  &   118 &      266  &     279  &  22--57 & 22--34 \& 49.0--57 & --\\[+0.3ex] 
J094331.61$+$053131.4 &0.548494  &   169 &      369  &     168  &  19--58 & 19--34 \& 42.4--58 &19--23 \\[+0.3ex] 
J100902.07$+$071343.9 &0.227855  &    70 &      540  &     450  &   0--66 &  0--14 \& 33.8--66 &0--3 \\[+0.3ex] 
J113910.79$-$135043.6 &0.212259  &   172 &   $>$1000 &  $>$1000 &  10--81 &       15--81       &--   \\[+0.3ex] 
J113910.79$-$135043.6 &0.319255  &    47 &   $>$1000 &     996  &  50--83 &       53--83       &65--83   \\[+0.3ex] 
J124154.02$+$572107.3 &0.205267  &    26 &      428  &     223  &  11--55 & 11--29 \&  42--55  & 11--16  \\[+0.3ex] 
J155504.39$+$362847.9 &0.189201  &    43 &      277  &     158  &  33--50 &       33--50       & 33--36  \\[+0.3ex] 
J225357.74$+$160853.6 &0.153718  &    31 &      162  &      74  &  46--60 &       46--60       & 46--50  \\[+0.3ex] 
J225357.74$+$160853.6 &0.352787  &   342 &      192  &     184  &   1--26 &        1--26       &   --
\enddata 
 \tablenotetext{a}{The height about the disk.}
 \tablenotetext{b}{The velocity of the red side of the cone at the lowest value of the opening angle.}
 \tablenotetext{c}{The velocity of the blue side of the cone at the lowest value of the opening angle.}
 \tablenotetext{d}{The half opening angle constrained by geometric arguments only.}
 \tablenotetext{e}{The half opening angle constrained by geometric arguments and for velocites less than 1000~\kms.}
 \tablenotetext{f}{Half opening angles where accelerated outflows exist.}
\end{deluxetable*}

For the majority of the opening angle range, the outflow velocities
required to reproduce the observations would be decelerating as the
gas moves further away from the host galaxy. With the assumed velocity
cut of 1000~{\kms}, there still remains a large range of opening
angles that are valid (see Table~\ref{tab:model}). This would imply
that either active outflows exist, and at these large heights above
the disk, the gas is rapidly decelerating or the absorption is a
result of previously ejected gas that is potentially falling back onto
the galaxy.

A caveat of these models is that we have assumed that all of the gas
seen in absorption is a result of the outflow. If only some fraction
of the gas is associated with outflows, then the model velocities, and
where acceleration and deceleration occur, would be different and
likely are expressed as upper limits. However, we do not have any
evidence to counter this assumption. Thus, we find that accelerating
outflow gas can only occur over a very small range of opening angles
and most of the time the gas is found to be decelerating.

\section{AMR Cosmological Simulations}\label{sec:sim}

We use cosmological simulations to provide further insight into what
mechanisms are driving the observed {\OVI} velocity spread.  These
hydrodynamical simulations provide the theoretical means to fully
incorporate dynamical processes, such as accretion and outflows, in a
cosmological setting. We apply the method of quasar absorption lines
to the simulations to observe the {\OVI} absorption kinematics. Here
we analyze eight $z=1$ simulated galaxies to identify the possible
structures and mechanisms that give rise to the observed {\OVI} halo
gas kinematics.

\subsection{Description of The Simulations}

We analyzed $\Lambda$CDM cosmological simulations created using the
Eulerian Gasdynamics plus N--body Adaptive Refinement Tree (ART) code
\citep{kravtsov99,kravtsov03}. The zoom-in technique \citep{klypin01}
applied here allows us to resolve the formation of single galaxies
consistently in their full cosmological context.

We analyzed the VELA simulation suite \citep{ceverino14,zolotov15},
which were created to compliment the {\it HST} CANDELS survey
\citep{barro13,barro14}.  The hydrodynamic code used to simulate these
galaxies incorporates prescriptions for star formation, stellar
feedback, supernovae type II and Ia metal enrichment, radiation
pressure, self-consistent advection of metals, and
metallicity-dependent cooling and photo-ionization heating due to a
cosmological ultraviolet background.  Our simulations have a feedback
model, named RadPre\_LS\_IR \citep{ceverino14}, that differs from
previous studies \citep{zolotov15}. This model includes radiation
pressure from infrared photons, as well as
photoheating/photoionization around young and massive stars. Further
details regarding the various models included in these simulations can
be found in \citep{ceverino09} and \citep{ceverino14}.

These simulations resulted in a maximum spatial resolution of 17~pc, a
dark matter particle mass of 8$\times$10$^4$~M$_{\odot}$, and a
minimum stellar particle mass of 10$^3$~M$_{\odot}$.  The high
resolution implemented in the VELA simulations allow us to resolve the
regime in which stellar feedback overcomes the radiative cooling
\citep{ceverino09}, which results in naturally produced galactic scale
outflows \citep{ceverino10,ceverino16}. Thus, galaxy formation
proceeds in a more realistic way through a combination of cold flow
accretion, mergers, and galaxy outflows.

Here we select a subsample of the VELA galaxies which a) were evolved
to the lowest redshift of $z=1$, b) did not experience a major merger
near $z=1$, and c) have a virial mass range between
log~M$_{vir}=11.3-12$ (see Table~\ref{tab:sims} for halo virial
quantities). The selection resulted in 8 galaxies having an average
log~M$_{vir}=11.7\pm0.2$~M$_{\odot}$ and
log~M$_{*}=10.5\pm0.3$~M$_{\odot}$.

\begin{deluxetable}{lccc}[!ht]
\tabletypesize{\scriptsize}
\tablecaption{Properties of $z=1$ VELA galaxies \label{tab:sims}}
\tablecolumns{4}
\tablewidth{0pt} 
\tablehead{
\colhead{VELA}&
\colhead{log($M_{vir}/$M$_{\odot}$)} &
\colhead{log($M_{*}/$M$_{\odot}$)}&
\colhead{$R_{vir}$} \\
\colhead{Galaxy}&
\colhead{ } &
\colhead{ }&
\colhead{(kpc)}
}
\startdata 
21  &12.0 & 10.9  &151\\[+0.3ex] 
22  &11.8 & 10.7  &133\\[+0.3ex] 
23  &11.7 & 10.4  &118\\[+0.3ex] 
25  &11.5 & 10.2  &103\\[+0.3ex] 
26  &11.6 & 10.4  &112 \\[+0.3ex] 
27  &11.6 & 10.3  &110  \\[+0.3ex] 
28  &11.3 & 9.9   & 92  \\[+0.3ex] 
29  &12.0 & 10.6  &146  
\enddata 
\end{deluxetable}

\subsection{Simulated Spectra}

We employed the HARTRATE photo+collisional ionization code
\citep{churchill14} that is optimally designed to model optically thin
gas with no ionization structure.  For the vast majority of the CGM,
including the {\OVI} column densities and impact parameters studied
here, this is a safe assumption.  The consequence of not including any
optical depth considerations is that HARTRATE may under-predict ions
that typically reside in optically thick conditions (such as {\MgII}),
however, this is not a concern here since this assumption only breaks
down close to central galaxies or near satellite galaxies. A proper
treatment of the radiation field would require computationally
intensive full radiative transfer computations, which is beyond the
scope of this work.

In summary, HARTRATE incorporates photo-ionization, direct collisional
ionization, Auger ionization, excitation-autoionization,
photo-recombination, high/low temperature dielectronic recombination,
charge transfer ionization by H$^{+}$, and charge transfer
recombination by H$^0$ and He$^0$.  HARTRATE uses solar abundance mass
fractions \citep{draine11,asplund09}, a \citet{haardt12} ionizing
spectrum is used for the ultraviolet background and assumes ionization
equilibrium. The cosmological simulations provides HARTRATE with the
hydrogen number density, kinetic temperature and the gas metallicity
(i.e., supernovae type II and Ia yields). The outputs from HARTRATE
include the electron density, the ionization and recombination rate
coefficients, ionization fractions and the number densities for all
ionic species up to zinc. The software has been applied successfully
in previous works \citep{kacprzak12b,churchill12,churchill15} and see
\citet{churchill14} for details on the code and its successful
comparisons to Cloudy.

The methodology of producing mock observations of quasar sightlines
through the cosmological simulations is described in detail in
\citet{churchill15} and \citet{vanderthesis}. The outputs from
HARTRATE are applied to Mockspec, which performed the mock quasar
absorption analysis.  Mockspec is publicly available in a GitHub
repository\footnote{https://github.com/jrvliet/mockspec}.  We ran
HARTRATE on a smaller box size of 6~R$_{vir}$ along a side centered on
the dark matter halo of the host galaxy and drew 1000 lines of sight
within a maximum impact parameter of 1.5~R$_{vir}$.
 
Absorption spectra with the instrumental and noise characteristics are
generated assuming each cell gives rise to a Voigt profile at its line
of sight redshift.  The mock quasar sightline is then objectively
analyzed for absorption above the equivalent width threshold of
0.02~{\AA}, which corresponds to $\log N({\OVI}) = 13.55$~{\cmsq} for
$b=10$~{\kms}.  The optical depth weighted median redshifts,
rest--frame equivalent widths and velocity widths and column densities
are then measured from the spectra \citep[see][]{cv01}.  The velocity
zero point of the simulated absorption lines is set to the line of
sight velocity of the simulated galaxy (center of mass of the
stars). For this analysis, all the eight simulated galaxies are
analyzed with the disk appearing edge-on to the observer.  The galaxy
inclination is determined relative to the angular moment vector of
cold gas ($T<10^4$~K) within 1/10 of R$_{vir}$. The systemic velocity
of the galaxy is determined by the dark matter particles within the
halo virial radius.

To examine the spatial and kinematic properties of gas giving rise to
{\OVI} absorption, we identify {\OVI} absorbing gas cells along each
sightline as those which contribute to detected absorption in the
simulated spectra. The gas cells along the sightline are sorted into
decreasing column density and the lowest are systematically removed
until the noiseless spectrum created by the remaining cells has an
equivalent width that is 95\% of the equivalent width of the original
spectrum.

\begin{figure}
\begin{center}
\includegraphics[angle=0,scale=0.44]{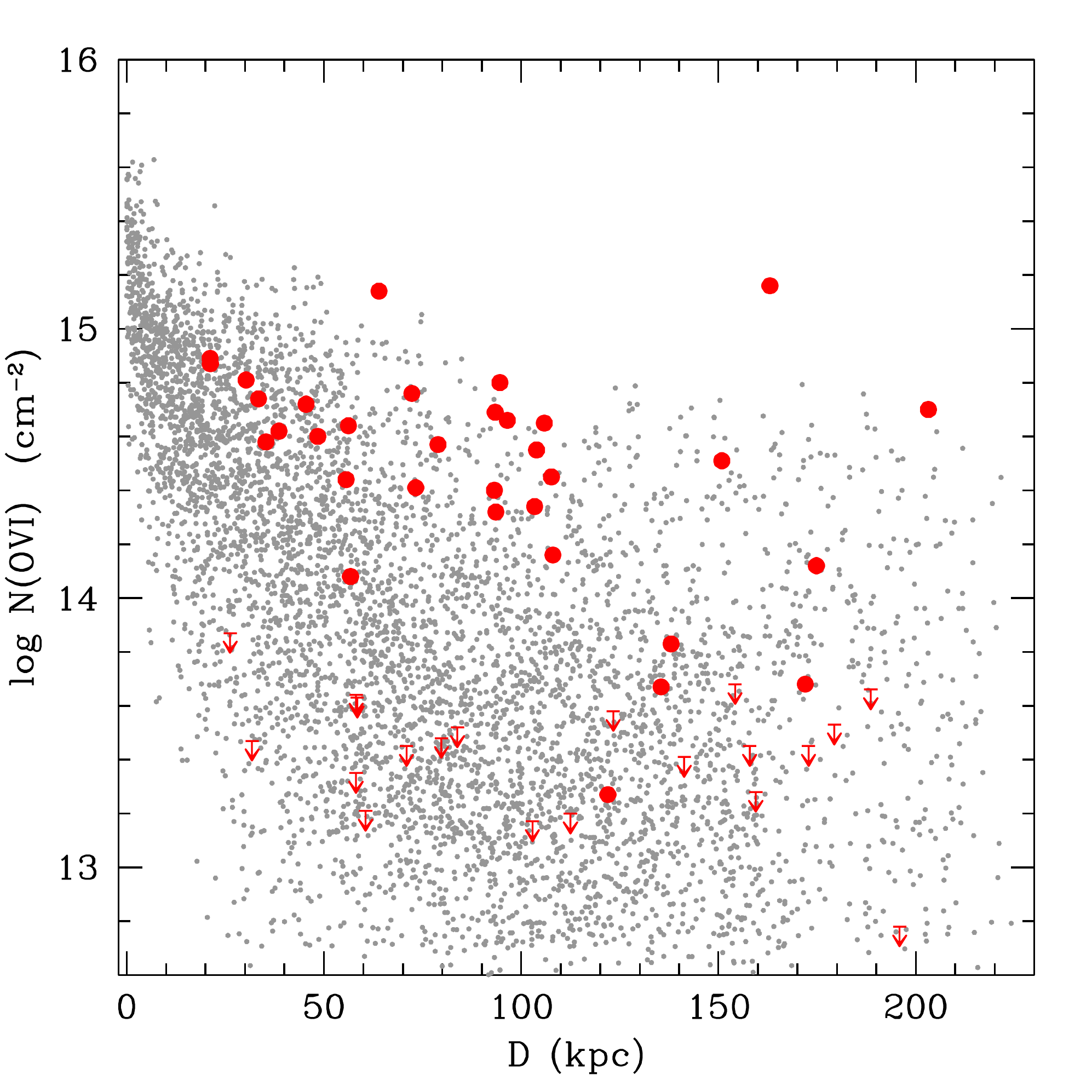}
\caption[angle=0]{{\OVI} column densities are shown as a function of
  impact parameter. Red points are observations taken from
  \citet{kacprzak15}. Grey points are from mock sightlines around 8
  simulated galaxies as described in the text.}
\label{fig:simsEW}
\end{center}
\end{figure}
\begin{figure*}
\begin{center}
\includegraphics[width=\linewidth]{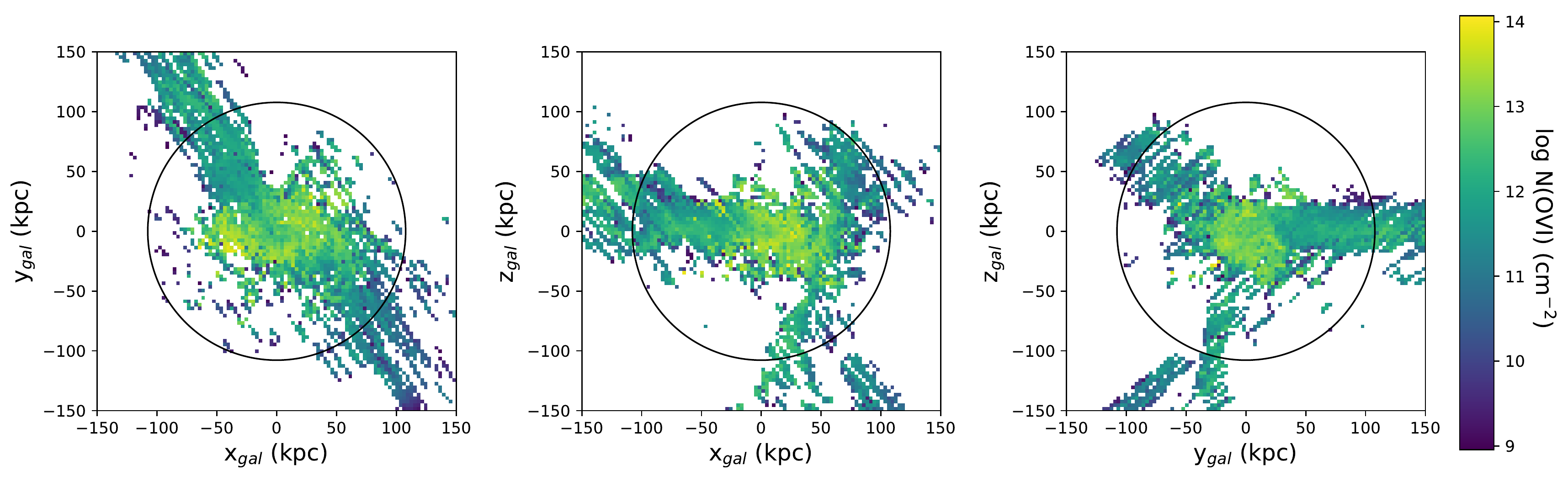}
\caption[angle=0]{Median {\OVI} column density spatial distribution
  located along sightlines drawn through an example simulated galaxy.
  The coordinate system is defined so the disk lies in the $xy$-plane
  with the angular moment vector of cold gas along the positive $z$
  axis.  {\OVI} absorption cells are shown for those that contribute
  to the absorption profiles (see text for methodology). The black
  circle shows the virial radius. Note 2--3 large filament structures
  that extend beyond 150~kpc around the galaxy. The {\OVI} within the
  central 40~kpc of the galaxy has a roughly spherical distribution.}
\label{fig:NOVI}
\end{center}
\end{figure*}

\subsection{Results Derived From Simulations}\label{sec:sims}

In Figure~\ref{fig:simsEW}, we show the {\OVI} column density
distribution from the simulations (grey) and from the observations of
\citet{kacprzak15} (red).  We note an anti-correlation for both the
observations and simulations between the column density and impact
parameter. There is overlap between the simulated and observe column
densities, while also being consistent with previous works using
simulations
\citep{hummels13,ford16,liang16,oppenheimer16,gutcke17,suresh17}. Although
the simulations shows a larger degree of scatter, which coud be driven
by galaxy inclination, etc., they can still provide useful insight
into the kinematics driving the existence of {\OVI} systems.  We will
explore this scatter and offsets between observations and simulations
in an upcoming paper.

Figure~\ref{fig:NOVI} shows the median {\OVI} column density
distribution for sightlines through the simulations for a single
example galaxy of VELA 27.  Only the cells contributing the the {\OVI}
absorption (as described in the previous section) are shown.  The
coordinate system for the example galaxy is defined so the disk lies
in the $xy$-plane with the angular moment vector of cold gas along the
positive $z$ axis. The black circle indicates the virial radius.  A
somewhat spherical {\OVI} halo is present within $\sim$40-50~kpc of
the galaxy center and has almost unity covering fraction. This
spherical halo around the host galaxy has column densities ranging
between log~$N$({\OVI})$=12.5-14$.

Beyond 50~kpc are possibly three thick filaments responsible for
producing the high impact parameter absorption with column densities
decreasing to log~$N$({\OVI})$=12.5-11$. These two features, halos and
streams, are seen in all of our simulated galaxies. Note that in this
particular example galaxy that the filaments are not co-planer and
tend to be in different locations for all galaxies. We will explore
the spatial distribution of {\OVI} in an upcoming paper. Next we
examine whether these structures in the simulations are able to
reproduce the typical absorption profiles and kinematics seen in our
observations.

\begin{figure*}
\begin{center}
\includegraphics[width=\linewidth]{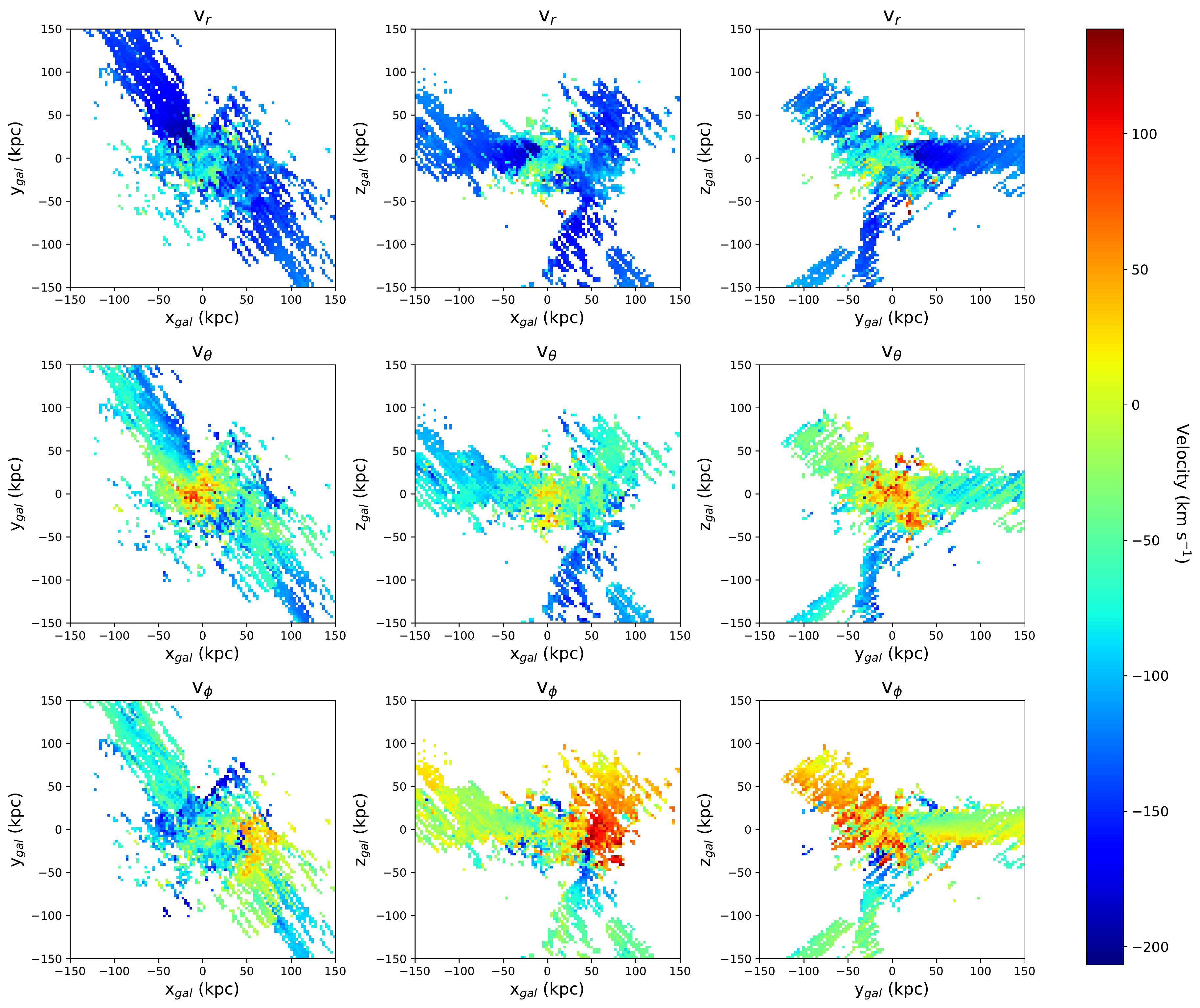}
\caption[angle=0]{The {\OVI} spatial distribution located along
  sightlines drawn through an example simulated galaxy.  The
  coordinate system is defined so the galaxy disk lies in the
  $xy$-plane with the angular moment vector of cold gas along the
  positive $z$ axis.  {\OVI} absorption cells are shown for those that
  contribute to the absorption profile (see text).  {\OVI} gas cells
  are color-coded by the median velocity along the projection in
  spatial coordinates (Top) $v_r$, (Middle) $v_{\theta}$ and (Bottom)
  $v_{\phi}$. For $v_{\theta}$, positive velocities indicates gas
  co-rotation with the same direction as the galaxy, which occurs for
  {\OVI} gas within 25~kpc of this example galaxy. Note both the
  significant radial inflow along the filaments and the co-rotating
  {\OVI} near the galaxy disk.}
\label{fig:velOVI}
\end{center}
\end{figure*}
Figure~\ref{fig:velOVI} shows the same {\OVI} gas cells contributing
to the absorption profiles as seen in Figure~\ref{fig:NOVI}, but now
color-coded as a function of velocity in spherical coordinates $v_r$,
$v_{\theta}$ and $v_{\phi}$.  Here the median velocity of all the
{\OVI} cells contributing to the absorption along each projection of
the sightlines are shown.

The top panel has the radial velocity component showing what speeds
the {\OVI} gas is traveling directly away or towards the center of the
galaxy. It can be clearly seen that there is significant radial inflow
towards the galaxy center along the filament structures. In this
particular example, the inflowing gas appears to have a roughly
constant velocity ranging between $-$150 to $-$200~{\kms}, with
potentially an increase towards the galaxy center.  The central part
of the galaxy halo has a component that exhibits slower inflow
velocities of 0--100~{\kms} that sits both near and outside of the
filaments. Most of the gas near the galaxy averages along the line of
sight is close to the systemic velocity. We see only a few gas cells
in this example that have positive, radially outflowing, velocities
ranging from 0--100~{\kms}.

The middle panel shows $v_{\theta}$, which is the rotation velocity,
where gas co-rotating with the galaxy has positive speeds and gas
counter-rotating with the galaxy is galaxy has negative velocities. In
the inner 50~kpc, the gas is rotating in the same direction as the
galaxy having velocities between 50--100~{\kms}. This co-rotating gas
appears to be in the same plane as the edge-on disk galaxy, suggesting
some connection between the {\OVI} and disk gas. There is also some
gas within 50~kpc that is near the systemic velocity and some
counter-rotating with speeds $<50$~{\kms}. Beyond 50~kpc, most of the
gas showing little sign of rotation while some gas is counter-rotating
with a range of velocities from 0 to $-$150~{\kms}. The dominant
velocity component outside of 50~kpc is the radial component.

The last panel shows $v_{\phi}$, which is the rate of change of the
angle between the vector to the gas cell and the $z-$axis, which is
aligned with the galaxy's angular momentum vector. In the central
region, we see positive velocities which decrease closer to systemic
velocity with increasing impact parameter. There are some negative
velocities out in the filaments as well.

We next explore the 8 simulations in a statistical sense in order to
determine general kinematic trends and origins of the {\OVI} CGM.

\begin{figure}
\begin{center}
\includegraphics[angle=0,scale=0.45]{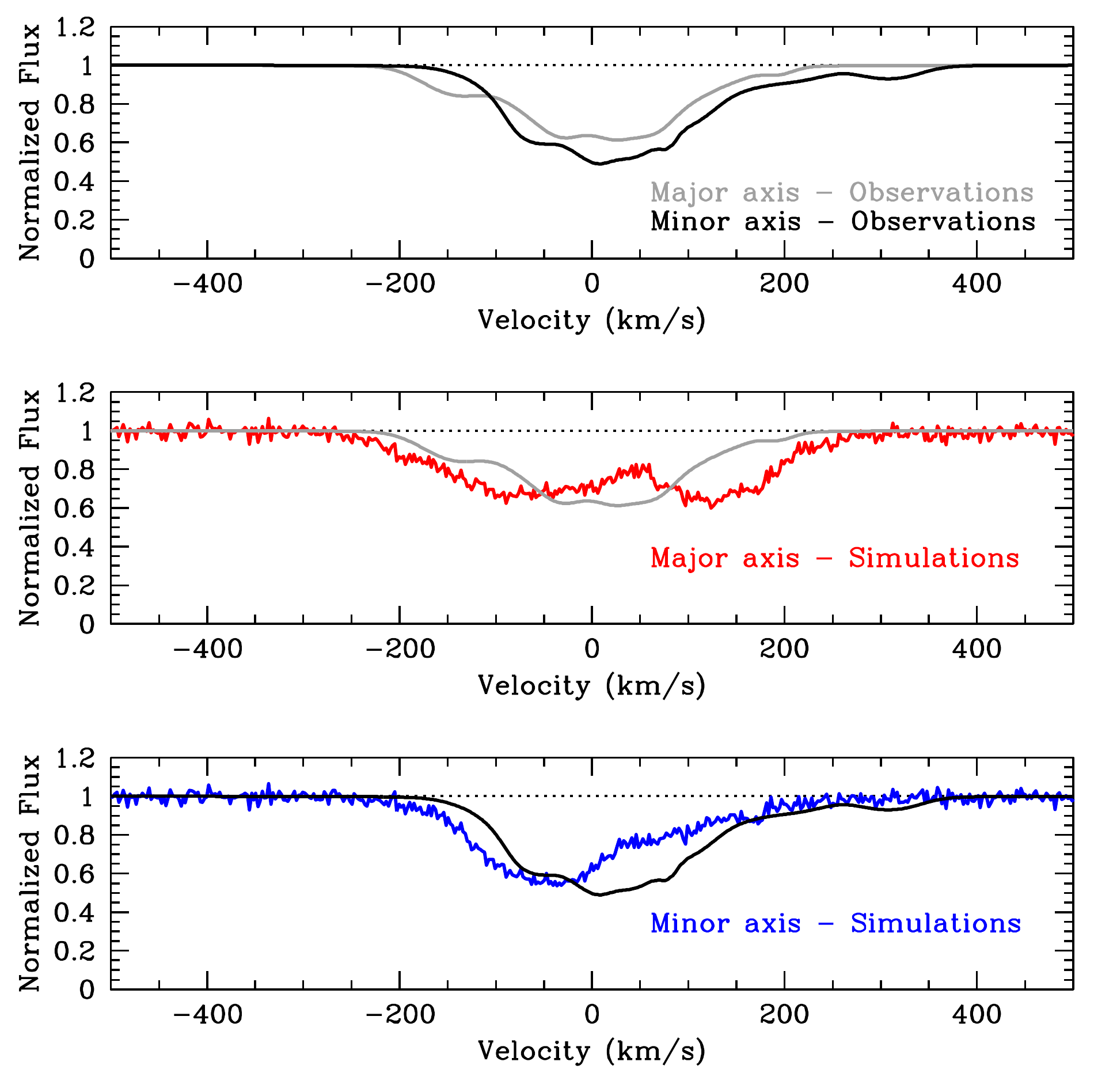}
\caption[angle=0]{(Top) Average observational {\OVI} spectra of 10
  sightlines along both the major (grey) and minor (black) axes of our
  sample (see Section~3). Both absorption profiles have similar line
  shapes and kinematics and are centered near the galaxy systemic
  velocity. (Middle) Observational {\OVI} major axis average spectrum
  shown along with the average spectrum from the simulations for major
  axis {\OVI} absorption. The simulated galaxy major axis is defined
  as having an azimuthal angle less than 30 degrees with absorption
  systems with equivalent widths of $>0.2$~{\AA}. The simulated
  spectra were computed using all sightlines along the major axis of
  all 8 simulated galaxies. (Bottom) Observational {\OVI} minor axis
  average spectrum shown along with the average spectrum from the
  simulations for minor axis gas having an azimuthal angle greater
  than 40 degrees for absorption systems having equivalent widths of
  $>0.2$~{\AA}. Note the similar optical depths between the
  observations and simulations, while they differ in their kinematic
  profiles.}
\label{fig:simsspec}
\end{center}
\end{figure}
\begin{figure*}
\begin{center}
\includegraphics[width=\linewidth]{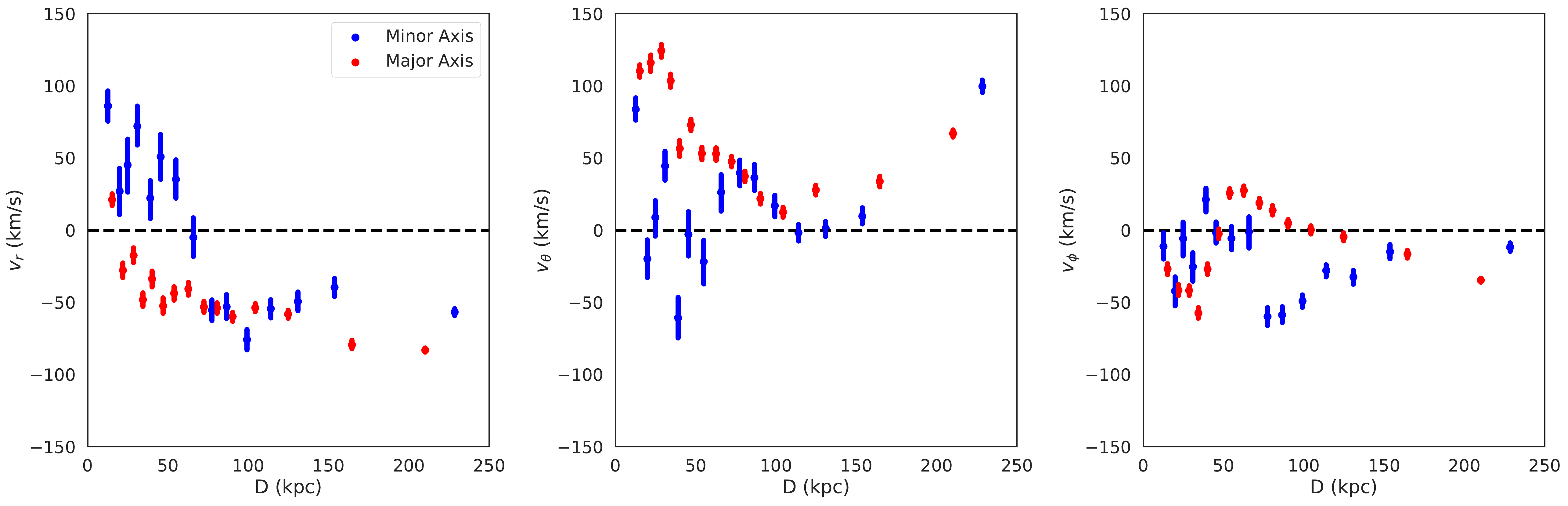}
\caption[angle=0]{Median {\OVI} velocities averaged over the eight
  simulations shown for the major axis and minor axis.  Major axis gas
  is defined as having an azimuthal angle less than 30 degrees while
  minor axis gas has an azimuthal angle greater than 40 degrees.  The
  first panel shows the radial velocity for major and minor axis gas
  in red and blue respectively along with the standard error in the
  mean. The middle panel shows the rotational angular velocity where
  positive velocities indicate {\OVI} gas rotating in the same
  direction of the galaxy. The last panel shows the polar velocity,
  which is the rate of change of the angle between the vector to the
  cell and the z-axis, which is aligned with the galaxy's angular
  momentum vector.}
\label{fig:avgvel}
\end{center}
\end{figure*}

The top panel of Figure~\ref{fig:simsspec} shows the mean stacked
Voigt profile fits to the {\OVI} that is located along the galaxy
major ($\Phi<25^{\circ}$) and minor ($\Phi<33^{\circ}$) axes for our
observations. Note both have similar kinematic shape and are centered
near the galaxy systemic velocity. The major axis gas is offset by
2.5~{\kms} from the galaxy systemic velocity while the minor axis gas
is offset by 28.0~{\kms} from the galaxy systemic velocity. This
implies that there are no strong kinematic signatures present if
outflows and accretion are traced by {\OVI} gas, or outflow and
accretion signatures could be hidden by a larger diffuse collection of
{\OVI} within the halo at similar velocities.

To compare our observations to the simulations, we select all
absorption systems from the 8 simulated galaxies that have an
equivalent width larger than 0.2~{\AA}, which is roughly the
observational limit of our sample. For the simulations, major axis gas
is defined as having an azimuthal angle less than 30 degrees, while
minor axis gas has an azimuthal angle greater than 40 degrees. These
absorption systems were then combined to provide the mean stacked
spectra shown in Figure~\ref{fig:simsspec}. We note that the optical
depths and the velocity spread between the simulations and
observations are similar, with some differences with the kinematic
shape of the profile.

The {\OVI} found near the major axis in the simulations exhibits a
possible bimodal distribution with bulk of the absorption residing
near $100-125$~{\kms} of each side of the galaxy systemic
velocity. This signature is reminiscent of co-rotation/accretion
predictions \citep{stewart11,stewart13,danovich15,stewart17}. {\OVI}
found near the minor axis in the simulations exhibits an offset of
$\sim50$~{\kms} from the galaxy systemic velocity but has a similar
velocity spread to the observations.  The simulated mean stacked
spectra do show some hints of kinematic structures, such as rotation
along the major axis, which does differ from our observations. This
could be due to only having 10 sight-lines from our observations, or
differences due to inclination angle effects.  We next examine the
typical {\OVI} velocities to determine what is driving the {\OVI}
kinematic distribution within the simulations.


Figure~\ref{fig:avgvel}, shows the median {\OVI} cell velocities
averaged over the eight simulations shown for gas along the major
(red) and minor (blue) axis. The first panel shows the radial velocity
component for major and minor axis gas along with the standard error
in the mean. Gas along the major axis of the galaxy appears to inflow
towards the galaxy at high velocities at high impact parameter and
slows to the galaxy systemic velocity as it approaches the galaxy
center. The largest deceleration occurs within 50~kpc, reducing in
speed from $-50$ to $0$~{\kms}. Thus, both Figures~\ref{fig:velOVI}
and \ref{fig:avgvel} indicate that {\OVI} gas does inflow along
filaments and decelerating as it approaches the galaxy.

On the other-hand, minor axis {\OVI} is outflowing out to about
50~kpc, then it decelerates and falls back towards the galaxy. The
minor axis gas has similar radial velocities as the major axis gas
beyond 75~kpc, which would make it difficult to identify the
difference between accreting and re-accreted gas. Thus, outflows
traced by {\OVI} only influence the CGM out to 50~kpc for a Milky Way-like
galaxy and recycled outflows, which are a common prediction from
simulations as an origin of {\OVI} gas, dominate at higher impact
parameters. This is consistent with the toy outflow models in
Section~\ref{sec:minor}, indicating that if the gas is originating
from outflows, the gas has to be decelerating and possibly falling
back to the galaxy. Furthermore, our minor axis observational sample
contains 3 galaxies with impact parameters less than 50~kpc. In those
three cases (J1241, J1555, J2253 $z_{gal}=$0.1537), the {\OVI} resides
to one side of the galaxy systemic velocity so it is possible that
those exhibit signatures of gas outflows.

The middle panel shows the rotational angular velocity, $v_{\theta}$,
where positive velocities indicate gas is rotating in the same
direction as the galaxy.  The major axis gas is rotating in the same
direction as the galaxy as it infalls towards the disk. The rotation
velocity component increase within 100~kpc and becomes the dominant
velocity component near the galaxy. Thus, we should see clear
signatures of co-rotation in our observations.  The minor axis gas may
be rotating in a similar direction within 25~kpc, but then scatters
around zero, showing little sign of following the direction of galaxy
rotation.

The last panel shows the polar velocity which is the rate of change of
the angle between the vector to the cell and the z-axis, which is
aligned with the galaxy's angular momentum vector.  This is the lowest
velocity component for the major axis gas, showing that this gas is
primarily infalling, co-rotating and not mixing very much azimuthally.
The minor axis gas has roughly zero polar velocity within 50~kpc and
beyond 125~kpc. Between 50 and 125~kpc, the gas begins to have
negative velocities. This occurs over the same impact parameter range
where the radial velocity of the minor axis gas transitions from
outflowing accelerating velocities to decelerating and accreting
velocities, indicating a change in the behavior of the kinematics of
minor axis {\OVI} gas. This is a signature of the {\OVI} returning
back to the disk-plane of the galaxy. Overall the dominant minor axis
velocity component is radial, be it outflowing or accreting.

\section{Discussion}\label{sec:discussion}

The amount of {\OVI} surrounding galaxies is significant and we are
just beginning to understand the role of {\OVI} in the CGM and its
origins.

\citet{nielsen17} attempted to address the origins of the {\OVI}
absorption by examining their kinematic profiles. The {\OVI}
absorption velocity spread is more extended than for {\MgII}
absorption, suggesting the two ions trace different parts of the
CGM. Furthermore, in contrast to {\MgII} that shows different
kinematics as a function of galaxy color, inclination and azimuthal
angle, {\OVI} is kinematically homogeneous regardless of galaxy
property. This is consistent with our results where, unlike {\MgII},
we do not find any clear kinematic signature of co-rotation/accretion
or signatures of definitive outflowing gas relative to the host
galaxy. {\OVI} found along the major axis of galaxies tends to span
the entire rotation curve of their host galaxy, with the average
{\OVI} major axis spectra centered at the galaxy systemic velocity
(only offset by 2.5~{\kms}) and spans from roughly
$\pm200$~{\kms}. Only one of the {\OVI} major axis systems could be
explained by a co-rotation model. Overall, roughly 50\% of the {\OVI}
optical depth can be found to either side of the galaxy rotation curve
with no preference for rotation direction. It is still plausible that
some of the {\OVI} could be rotating in the same direction as the
galaxy, we just have no way of differentiating that component relative
to the rest of the {\OVI}.

We further find that the {\OVI} along the minor axis of galaxies
does not show clear signs of co-rotation, with only three of ten
systems that have relative galaxy and gas kinematics that can be
modeled well with a co-rotation model. The remainder of the systems
have the bulk of the gas counter-rotating with respect to the
galaxy. Maybe this is not so surprising given that the gas is not
located in the plane of the disk, but off-axis co-rotation is still
common for {\MgII} absorbers \citep[e.g.,][]{kacprzak10}.

We further apply simple outflow models in an attempt to constrain the
probability of outflows driving the observed {\OVI}
kinematics. Interestingly, we find that accelerating outflows can only
occur when opening angles are small.  The remainder of the parameter
space has the {\OVI} decelerating and falling back on the galaxy.
This could be why in the stacked minor axis {\OVI} profiles only have
a systematic offset of 28~{\kms} from the galaxy systemic
velocity. Again, this would be kinematically different compared to
what is seen for {\MgII} where, over a similar impact parameter range,
{\MgII} gas tends to have accelerated flows
\citep{bouche12,bordoloi14,schroetter16}. However, these observed
      {\OVI} kinematics are consistent with simulations having
      predicted that a possible origin of {\OVI} is from ancient
      outflows, which would eventually fall back to the galaxy
      \citep[e.g.,][]{ford14,ford16,oppenheimer16}. So it is possible
      that we are seeing the kinematic signatures of the gas recycling
      from ancient outflows.

Our simulations show that {\OVI} can be found in filamentary
structures and within outflow winds as seen in
Figure~\ref{fig:NOVI}. The {\OVI} has a radial velocity flow towards
the galaxy starting at $-$80~{\kms} at 200~kpc and reduces in speed as
it approaches the galaxy along with major axis (see
Figure~\ref{fig:avgvel}). This the rotational speed of the infalling
gas also increases as it approaches the galaxy and shows little sign
of azimuthal mixing as indicated by the low polar velocities.  We find
that minor axis {\OVI} outflows of a modest velocity 50~{\kms} occur
within the first 50~kpc, then decelerate and begin to fall back onto
the galaxy (as indicated by the $-50$~{\kms} polar velocities). These
gas flows appear quite obvious within the simulations, but the
simulations contain a wealth of information and 1000s of
lines-of-sight, so we typically show velocity medians and median
column densities, but this is not how we observe {\OVI} in
reality. What we normally observe is integrated velocities and optical
depths, which are quite different to median values.

\begin{figure}
\begin{center}
\includegraphics[width=\linewidth]{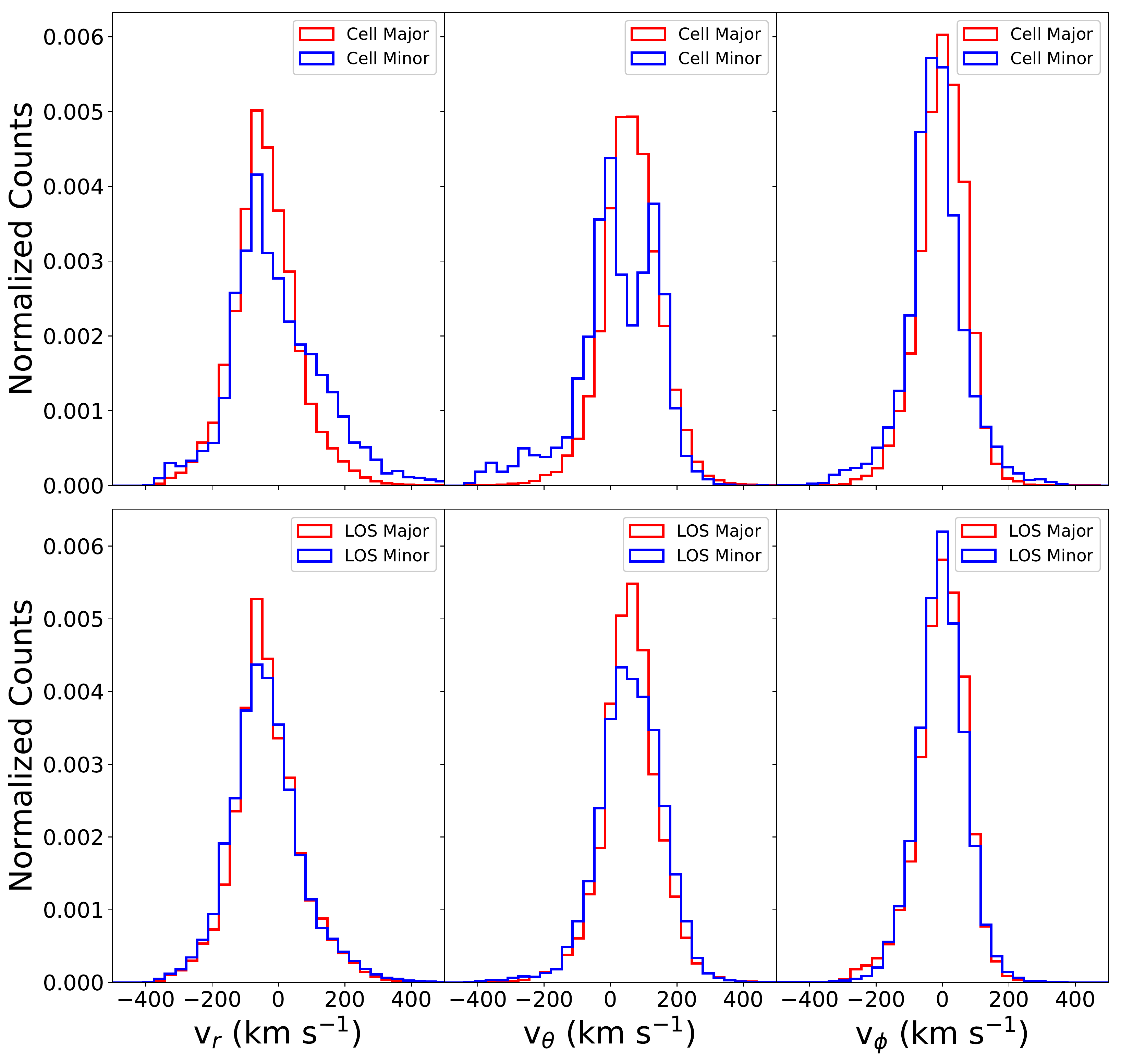}
\caption[angle=0]{ Histograms of the radial and rotational velocities
  from the eight simulations. (Top) We select gas cells within a cone
  of an opening angle $\pm$30 degrees around the major axis and
  $\pm$50 degrees around the minor axes showing the velocity histogram
  of gas that is likely infalling and outflowing. We choose these
  regions in order to select gas likely only associated with gas
  flows. (Bottom) Histogram of the velocities for all the gas cells
  along the quasar sightlines through the entire halo, selecting all
  gas though the halo showing the full range of velocities being
  intercepted. Note that significant kinematic features become lost
  and that there is a lot of gas at similar velocities both align
  along the major and minor axis.}
\label{fig:relvel}
\end{center}
\end{figure}

In Figure~\ref{fig:relvel}, we show the histograms of the radial,
rotational and polar velocities from the eight simulations. We define
two sets of data. In the top panel, we select {\OVI} gas cells within
a cone of a half-opening angle 30 degrees around the major axis and 40
degrees half-opening angle around minor axis, over all impact
parameters and show the velocity histogram of gas.  Both major and
minor axis gas peak at negative radial velocities since major axis gas
is flowing along filaments and the minor axis gas in falling back to
the galaxy, with some additional power at positive velocities for the
minor axis outflowing gas. For rotational velocity the major axis gas
peaks at positive velocities since it is rotating in the same
direction of the galaxy, while minor axis gas has a bimodal
distribution exhibiting both co- and counter-rotating velocities. The
major axis gas also exhibits a peak at systemic polar velocity while
minor axis gas peaks at negative velocities indicating gas can be
accreting back onto the galaxy.

In the bottom panel of Figure~\ref{fig:relvel}, we show a histogram of
velocities for all the {\OVI} gas cells along the quasar sightlines
through the entire galaxy halo. Note that significant kinematic
features become lost and major and minor axis gas have a similar
velocity structure, which is what we see in our observations shown in
Figure~\ref{fig:simsspec}. The stronger radial outflow component
becomes lost along with the co- and counter-rotating gas. Both major
and minor axis gas have similar distributions in all velocity
components.  This implies that gas all along the quasar sightlines
through the entire halo conspires to line up in velocity, masking any
signatures of gas flows. So although it is likely that there is some
fraction of the observed {\OVI} that could be tracing accretion and
outflows, we are unable to quantify this with our observations.

Although \citet{kacprzak15} reported that the spatial {\OVI} azimuthal
dependence is a result of gas major axis-fed inflows/recycled gas and
minor axis-driven outflows, it is impossible to confirm this using the
kinematics of {\OVI} alone.  Furthermore, \citet{nielsen17} postulated
that the higher column densities found near the major and minor axes
of galaxies, as traced by {\MgII} absorption, my provide a shield such
that {\OVI} is not so easily further ionized as it would be at
intermediate azimuthal angles. This would naturally produce an
azimuthal dependence without {\OVI} being directly linked to outflows
and accretion. Disentangling these two ideas will require much more
investigation using multi-phase gas tracers.

Overall, although the simulations indicate that {\OVI} is present in
inflowing and outflowing gas, observationally it seems that {\OVI} is
not ideal to use as a kinematic tracer of gas flows within the
galaxy. It might be further complicated by the fact that the
simulations predict collisional ionized {\OVI} is dominant in the
central regions of halos (inside $0.2-0.3$~$R_{vir}$), while
photoionization is more significant at the outskirts around $R_{vir}$
\citep{roca-fabrega18}.  It is likely that {\OVI} is more indicative
of the thermal motions of the gas probing the temperature of the dark
matter halos \citep{oppenheimer16,roca-fabrega18} though this remains
highly debated. A halo mass dependence has been directly observed for
{\OVI} \citep{ng18,pointon17} leading credence to the thermal
temperature model.

\section{Conclusions}\label{sec:conclusion}

We have constructed a sub-sample from \citet{kacprzak15} of 20 {\OVI}
absorption systems (EW>0.1~{\AA}) associated with isolated galaxies
that have accurate spectroscopic redshifts and rotation curves
obtained from Keck/ESI.  Given the observed {\OVI} azimuthal angle
bimodality \citep{kacprzak15}, our sample is split into two azimuthal
angle bins described as major axis ($\Phi<25$ degrees) and minor axis
($\Phi>33$ degrees) samples.  Our results are summarized as follows:

\begin{enumerate}

\item The {\OVI} absorption found along the major axis (within
  $\Phi=25$ degrees) of their host galaxy does not show any
  significant correlation with galaxy rotation and {\OVI}
  kinematics. Only one system can be explained by simple
  rotation/accretion model. This is in contrast to co-rotation
  commonly observed for {\MgII} absorption systems. The {\OVI}
  absorption kinematics span the entire dynamical range of their host
  galaxies and have a relative velocity offset of only 2.5~{\kms} from
  the galaxy systemic velocity.

\item The {\OVI} found along the minor axis of galaxies ($\Phi>33$
  degrees) could be modeled by outflows. Simple models show that only
  over a small parameter space (with small opening angles) {\OVI}
  can be accelerating in outflows. The rest of the time the gas must
  be decelerating and being recycled, which is consistent with
  simulations. The absorption redshift has a velocity offset of
  28.0~{\kms} relative to the host galaxy systemic velocity.

\item 3-D visualization of our simulations shows that {\OVI} is
  contained in filaments and in a spherical halo of $\sim$50~kpc in
  size surrounding the host galaxy. This implies that we should see
  kinematic signatures of {\OVI} within the simulations.

\item The {\OVI} absorption-lines created from sightlines passing
  through the simulations along the major and minor axes have similar
  optical depths, velocity widths and have differ only in a kinematic
  shape. This difference is likely attributed to differences in galaxy
  properties such as inclination.

\item All {\OVI} identified in the simulated sightlines along the
  major axis have kinematics consistent with gas accretion along
  filaments, which decelerate as they approach the host
  galaxy. Infalling gas also rotates in the same direction of the
  galaxy, and increases in velocity as it approaches the galaxy. Thus
  {\OVI} can trace gas accretion.

\item All {\OVI} identified in the simulated sightlines along the
  minor show that outflows only have positive velocities within the
  inner 50-75~kpc where they eventually decelerate and fall back in
  towards at around $-$50~{\kms}.

\item The kinematic signatures in the simulations are quite clear when
  computing median velocities and column densities. However, when we
  compare these to apparent kinematic signatures integrated along
  lines of sight, we find that strong gas kinematic signatures are
  washed out due to existing velocity structure from all the different
  structures through the halo and the diffuse gas between them.
 
\end{enumerate}

Although we do not know the true origins of {\OVI}, it appears to not
serve as a useful kinematic indicator of ongoing gas accretion,
outflows or star-formation. Ions such as {\MgII}, {\SiII} and {\CaII}
have all indicated that they are better tracers of gas kinematics even
over the same {\HI} column density range as {\OVI}. \citet{ng18} and
\citet{pointon17} show clear evidence that {\OVI} is halo mass
dependent, efficiently probing the viral temperature of the halo as
predicted in the simulations
\citep{oppenheimer16,roca-fabrega18}. Although {\OVI} can trace
interesting phenomena within galaxy halos, this is masked by all the
diffuse gas found ubiquitously within the halos at velocities of
$\sim$$\pm200$~{\kms}. The interest in {\OVI} has increased in recent
years due to the ease it can be simulated in cosmological simulations,
and from {\it HST}/COS initiatives, but we must now turn more of our
efforts to simulating the cool CGM in order to place reasonable gas
physics constraints on galaxy growth and evolution.

\acknowledgments We thank Roberto Avila (STScI) for his help and
advice with modeling PSFs with ACS and WFC3. GGK acknowledges the
support of the Australian Research Council through the award of a
Future Fellowship (FT140100933). GGK and NMN acknowledges the support
of the Australian Research Council through a Discovery Project
DP170103470. CWC and JCC are supported by NASA through grants HST
GO-13398 from the Space Telescope Science Institute, which is operated
by the Association of Universities for Research in Astronomy, Inc.,
under NASA contract NAS5-26555. CWC and JCC are further supported by
NSF AST-1517816.  SM acknowledges support from the ERC grant
278594-GasAroundGalaxies. DC been funded by the ERC Advanced Grant,
STARLIGHT: Formation of the First Stars (project number 339177).  The
VELA simulations were performed at the National Energy Research
Scientific Computing Center (NERSC) at Lawrence Berkeley National
Laboratory, and at NASA Advanced Supercomputing (NAS) at NASA Ames
Research Center.  Most of the data presented here were obtained at the
W. M. Keck Observatory, which is operated as a scientific partnership
among the California Institute of Technology, the University of
California, and the National Aeronautics and Space Administration. The
Observatory was made possible by the generous financial support of the
W. M. Keck Foundation.  Observations were supported by Swinburne Keck
programs 2016A\_W056E, 2015B\_W018E, 2014A\_W178E and
2014B\_W018E. The authors wish to recognize and acknowledge the very
significant cultural role and reverence that the summit of Mauna Kea
has always had within the indigenous Hawaiian community.  We are most
fortunate to have the opportunity to conduct observations from this
mountain. Based on observations made with the NASA/ESA Hubble Space
Telescope, and obtained from the Hubble Legacy Archive, which is a
collaboration between the Space Telescope Science Institute
(STScI/NASA), the Space Telescope European Coordinating Facility
(ST-ECF/ESA) and the Canadian Astronomy Data Centre (CADC/NRC/CSA).





{\it Facilities:} \facility{Keck II (ESI)}
\facility{HST (COS, WFPC2, ACS, WFC3)}.

\newpage
\begin{appendix}
\section{Major axis sample}\label{sec:A}

Here we show 10 systems where the {\OVI} absorption is detected with
25~degrees of the galaxy major axis. We present the data used in this
analysis, which lncludes {\it HST}/COS {\OVI} absorption spectra, {\it
  HST} imaging of the quasar and galaxy field, along with the Keck/ESI
spectra for each galaxy to derive their rotation curves. We further
present a simple rotating disk model as described in
Section~\ref{sec:majorrot}.

\begin{figure*}[h]
\begin{center}
\includegraphics[scale=0.47]{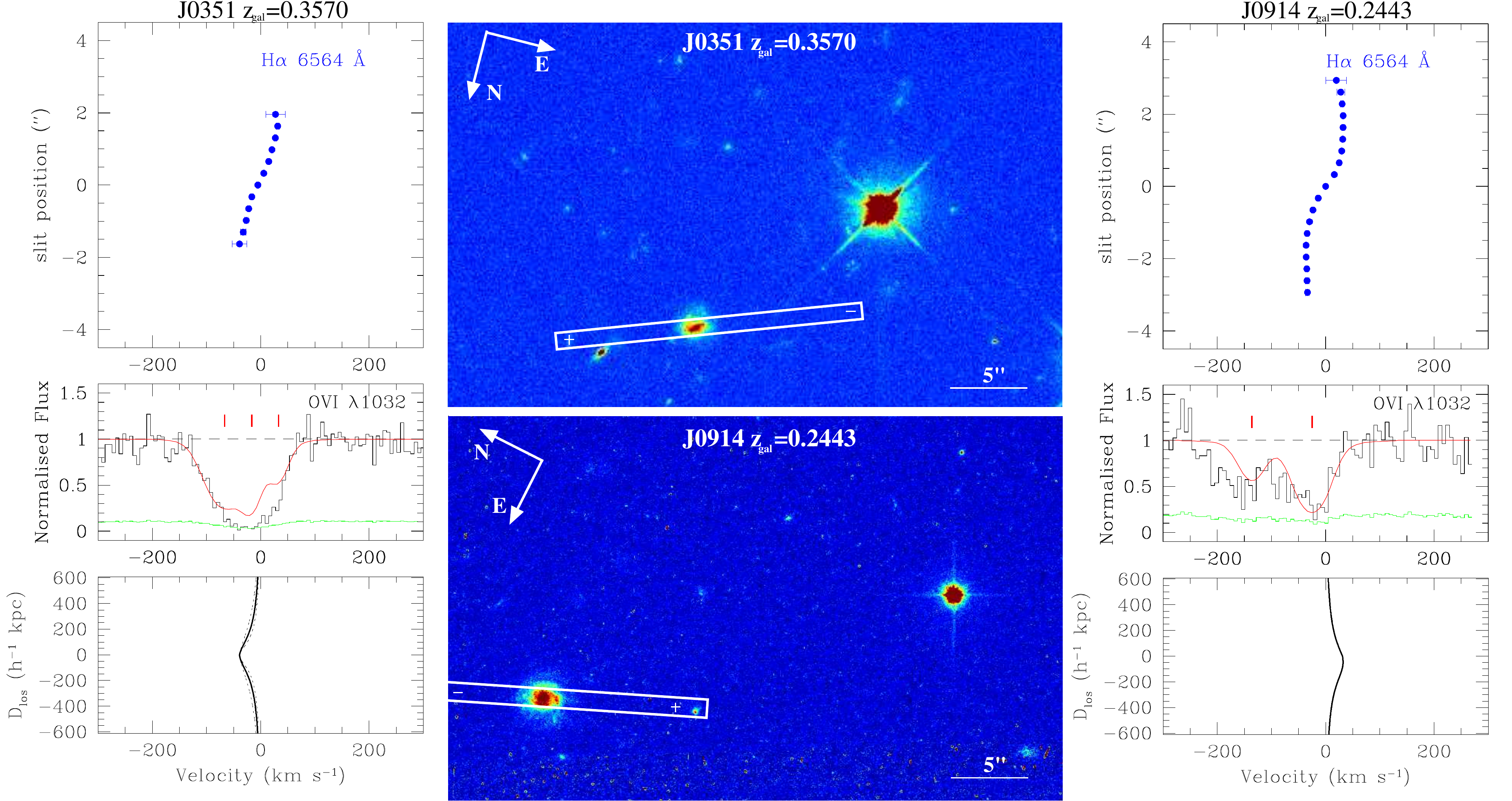}
\caption[angle=0]{Same as Figure~\ref{fig:moskine} -- {\it HST} images
  and galaxy rotation curves presented for two fields where the quasar
  sight-line aligns with the galaxy's major axis. (Top middle) A 45$''
  \times $25$''$ {\it HST} image of the quasar field J0351. The
  ESI/Keck slit is superimposed on the image over the targeted
  galaxy. The "$+$" and "$-$" on the slit indicate slit direction in
  positive and negative arcseconds where 0$''$ is defined at the
  galaxy center.  (Left) The $z=0.3570$ galaxy rotation curve and the
  {\it HST}/COS {\OVI} $\lambda$1031 absorption profile is shown with
  respect to the galaxy systemic velocity. The panel below the {\OVI}
  absorption is a simple disk rotation model computed using
  Equation~\ref{eq:kine}, which is a function of the galaxy rotation
  speed and orientation with respect to the quasar sight-line. The
  J0351 galaxy is rotating in the same direction as the absorption
  however, the velocity range covered by the model is not consistent
  with the entire range covered by the absorption profile.  (Bottom
  middle) Same as top middle except for the J0914 quasar field and for
  the targeted galaxy at $z = 0.2443$ (Right) Same as left except the
  $z = 0.2443$ in the J0914 quasar field. Note here that the {\OVI}
  absorption is consistent with being counter-rotating with respect to
  the galaxy and again, the model has insufficient velocities to
  account for all the absorption kinematics. In both cases
  disk-rotation does not reproduce the observed absorption
  velocities.}
\label{fig:A1}
\end{center}
\end{figure*}
\begin{figure*}
\begin{center}
\includegraphics[scale=0.47]{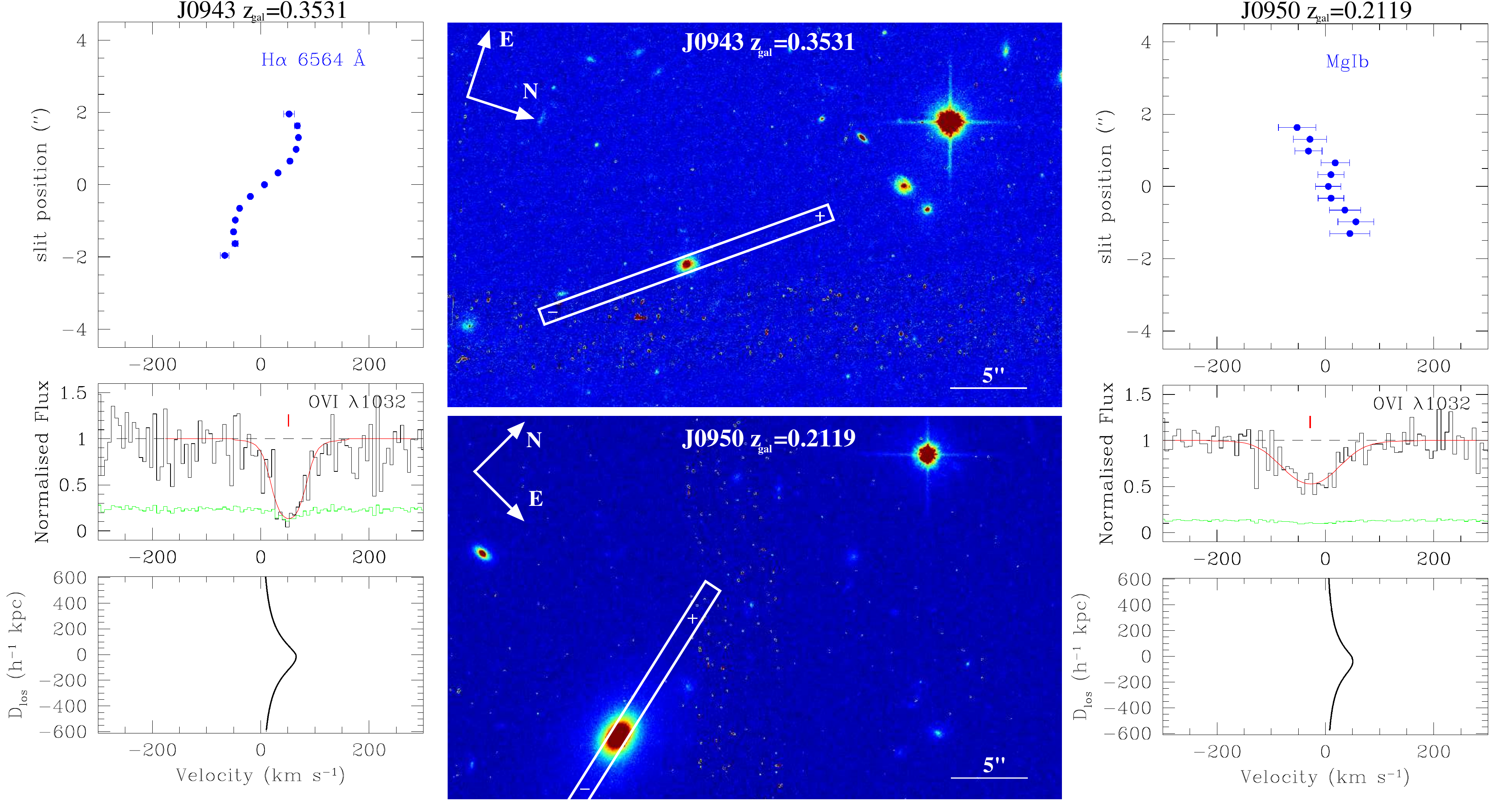}
\caption[angle=0]{Same as Figure~\ref{fig:A1} except for top-middle
  and left is for the J0943 field with the $z=0.3531$ galaxy and
  bottom-middle and right is for J0950 field with the $z=0.2119$
  galaxy. The J0943 $z=0.3531$ galaxy has a rotation velocity that
  matches the observed {\OVI} absorption kinematics. The J0950
  $z=0.2119$ galaxy is counter-rotating with respect to the bulk of
  the {\OVI} absorption.}
\label{fig:A2}
\end{center}
\end{figure*}
\begin{figure*}
\begin{center}
\includegraphics[scale=0.47]{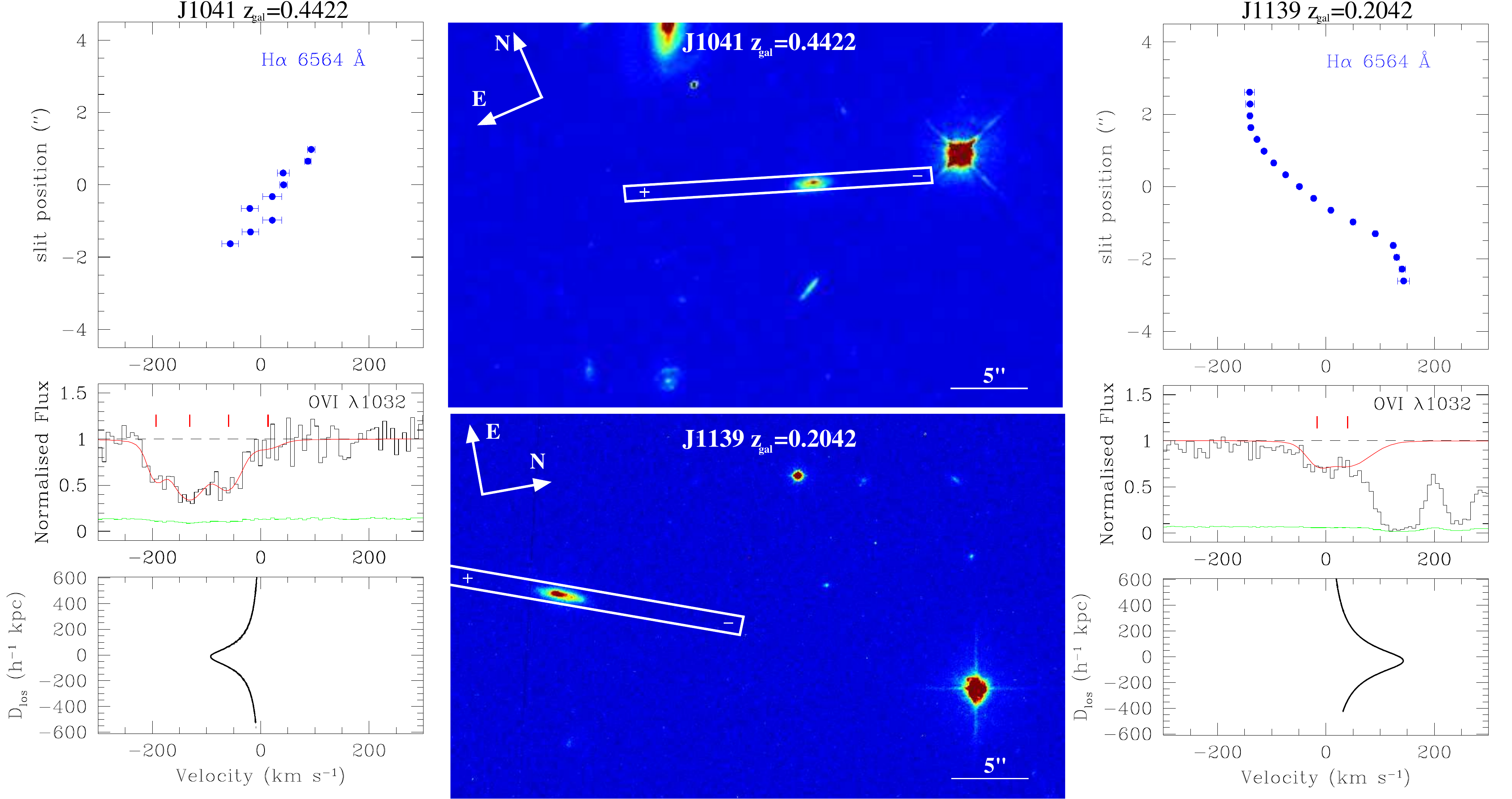}
\caption[angle=0]{Same as Figure~\ref{fig:A1} except for top-middle
  and left is for the J1041 field with the $z=0.4422$ galaxy and
  bottom-middle and right is for J1139 field with the $z=0.2042$
  galaxy. The J1041 $z=0.4422$ galaxy has a rotation velocity that
  matches the observed {\OVI} absorption kinematics. It slightly
  under-predicts the {\OVI} absorption velocity, yet this could be due
  to the not covering the full rotation curve with our
  observations.The J1139 $z=0.2042$ galaxy has a rotation velocity that
  matches the observed {\OVI} absorption kinematics.}
\label{fig:A3}
\end{center}
\end{figure*}
\begin{figure*}
\begin{center}
\includegraphics[scale=0.47]{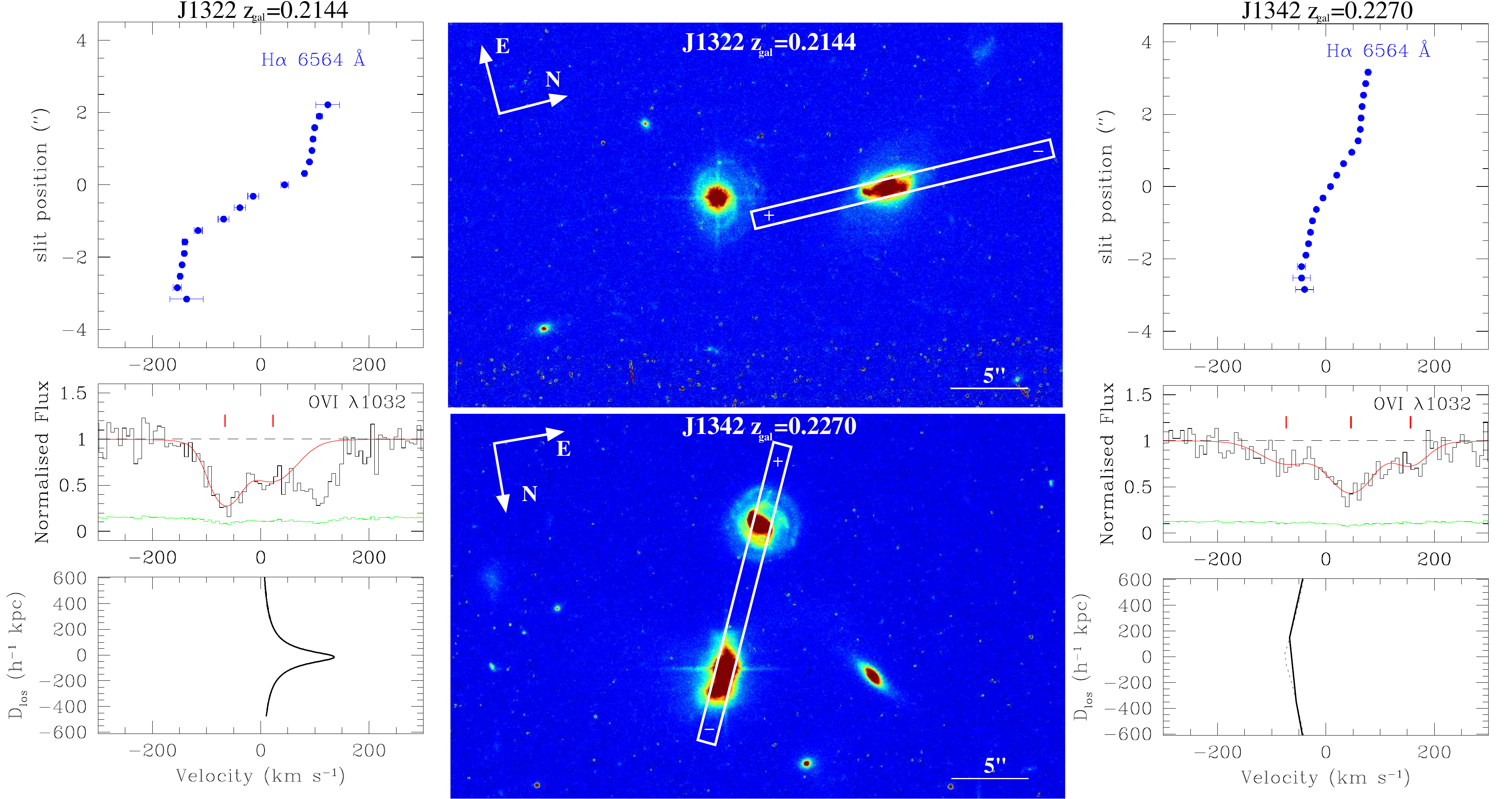}
\caption[angle=0]{Same as Figure~\ref{fig:A1} except for top-middle
  and left is for the J1322 field with the $z=0.2144$ galaxy and
  bottom-middle and right is for J1342 field with the $z=0.2270$
  galaxy. Both galaxies here are counter-rotating with respect to the
  bulk of the {\OVI} absorption.}
\label{fig:A4}
\end{center}
\end{figure*}
\begin{figure*}
\begin{center}
\includegraphics[scale=0.47]{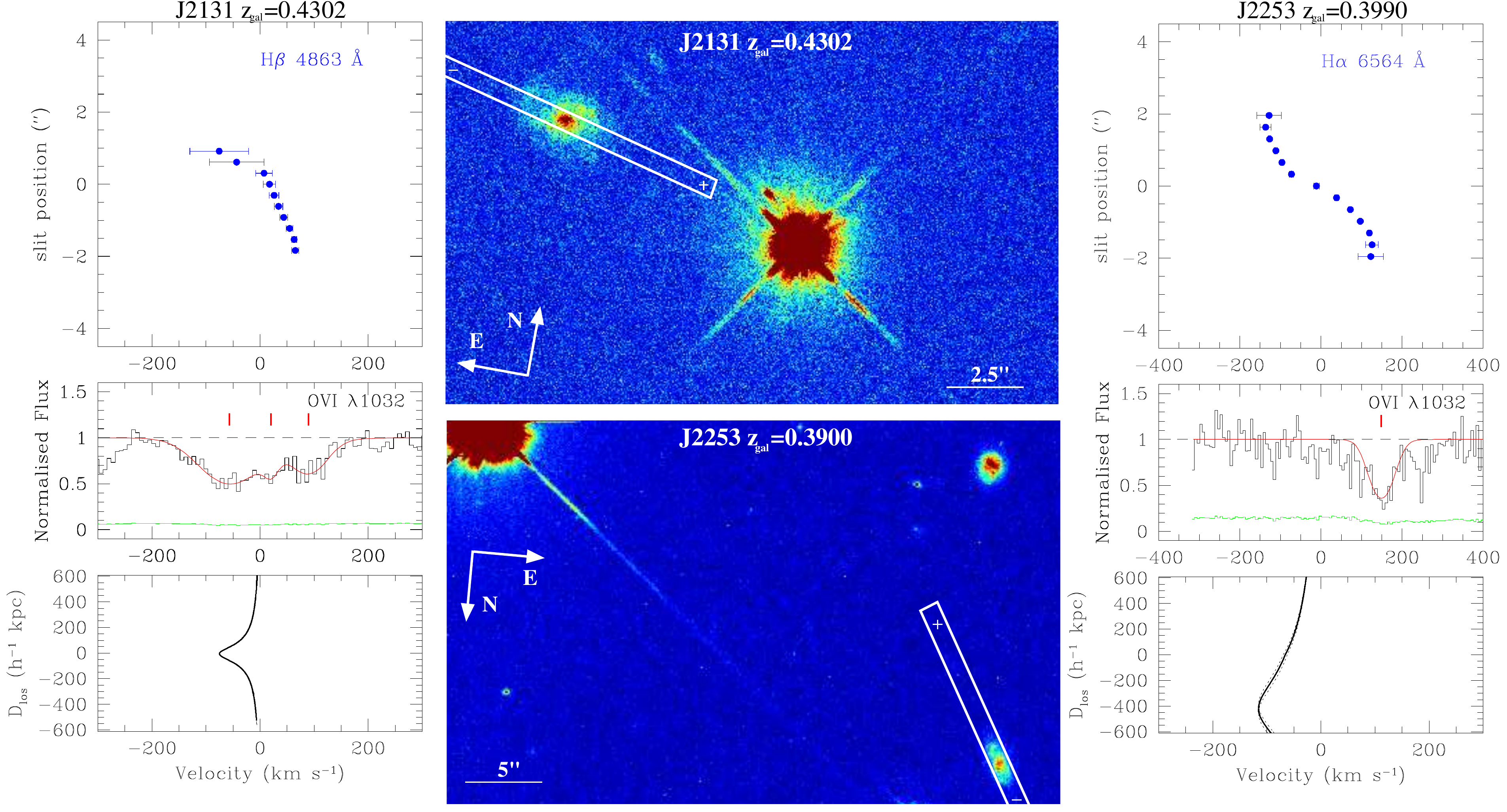}
\caption[angle=0]{Same as Figure~\ref{fig:A1} except for top-middle
  and left is for the J2131 field with the $z=0.4302$ galaxy and
  bottom-middle and right is for J2253 field with the $z=0.3900$
  galaxy. Both galaxies here are counter-rotating with respect to the
  bulk of the {\OVI} absorption.}
\label{fig:A5}
\end{center}
\end{figure*}
\clearpage

\section{Minor axis sample}\label{sec:B}

Here we show 10 systems where the {\OVI} absorption is detected
azimuthal angles of greater than 33~degrees as measured from the
galaxy major axis. We again present the data used in this analysis,
which lncludes {\it HST}/COS {\OVI} absorption spectra, {\it HST}
imaging of the quasar and galaxy field, along with the Keck/ESI
spectra for each galaxy to derive their rotation curves. Although this
is a minor axis sample, we stil present the simple rotating disk model
as described in Section~\ref{sec:majorrot}.

\begin{figure*}
\begin{center}
\includegraphics[scale=0.47]{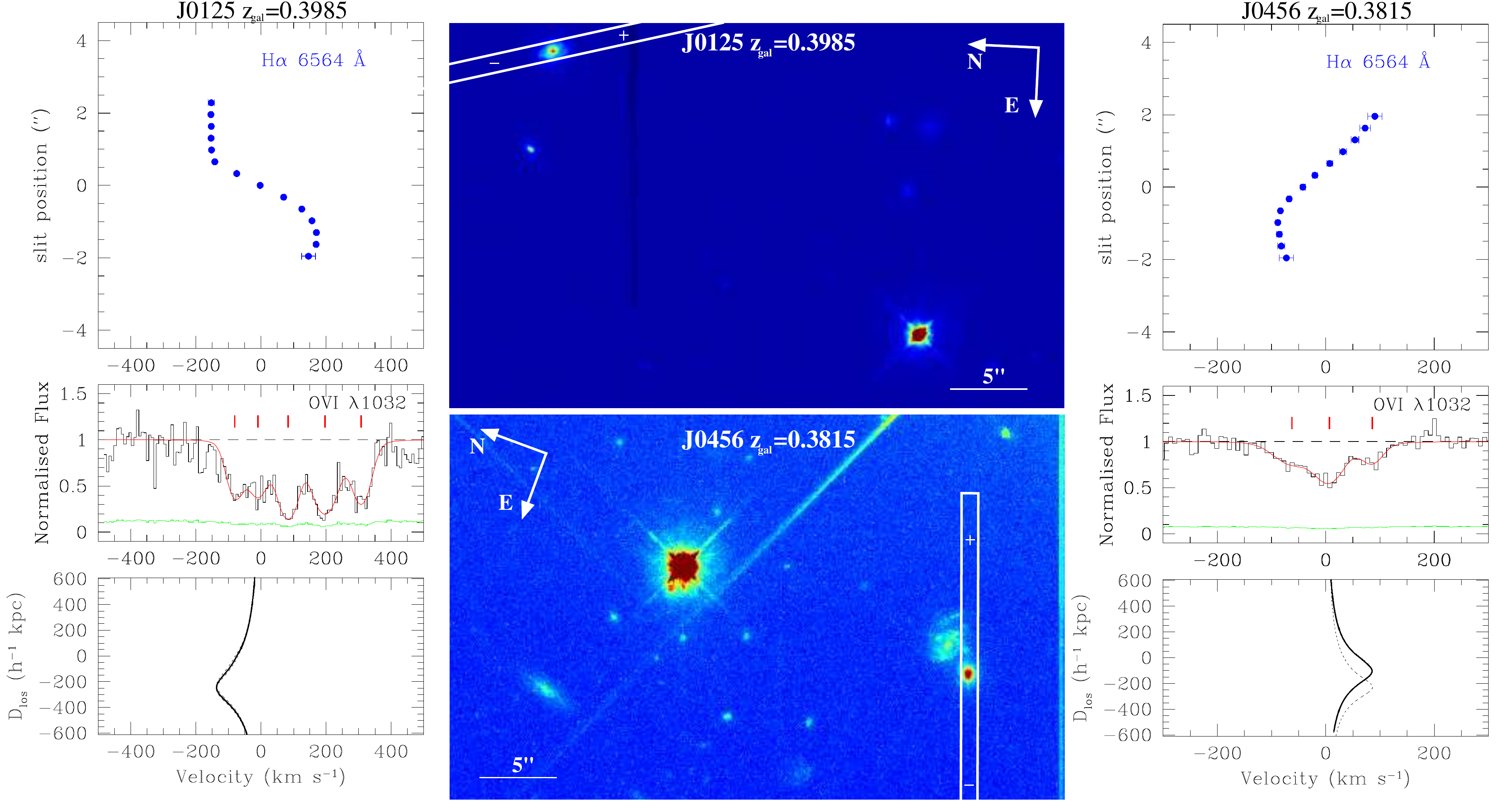}
\caption[angle=0]{Same as Figure~\ref{fig:A1} except for top-middle
  and left is for the J1025 field with the $z=0.3985$ galaxy and
  bottom-middle and right is for J0456 field with the $z=0.3815$
  galaxy. For the J1025 $z=0.3985$ galaxy, the {\OVI} has a large
  velocity spread that extends opposite to the rotational direction of
  the galaxy. The J0456 $z=0.3815$ galaxy is counter-rotating with
  respect to the bulk of the {\OVI} absorption.}
\label{fig:B1}
\end{center}
\end{figure*}
\begin{figure*}
\begin{center}
\includegraphics[scale=0.47]{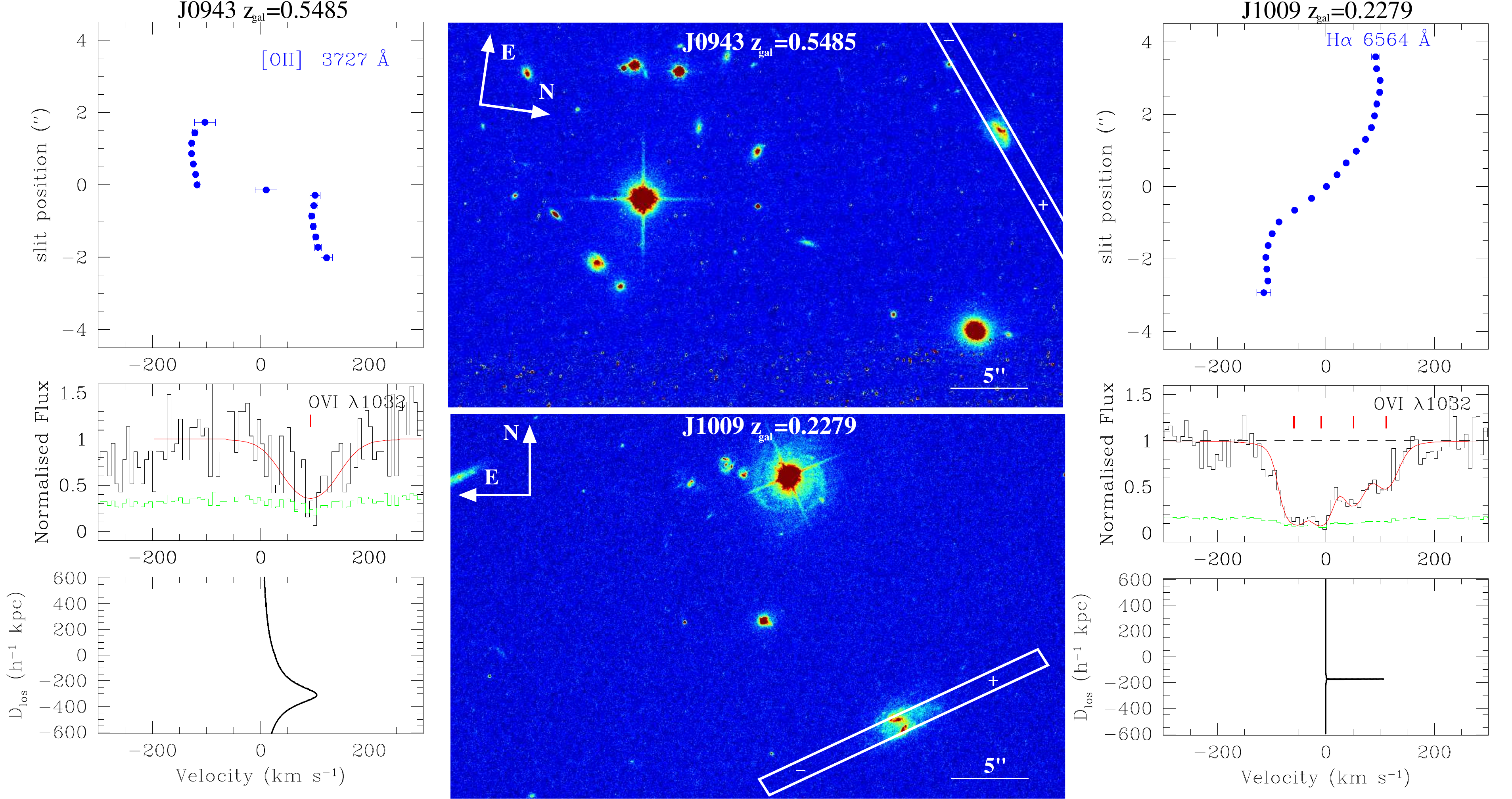}
\caption[angle=0]{Same as Figure~\ref{fig:A1} except for top-middle
  and left is for the J0943 field with the $z=0.5485$ galaxy and
  bottom-middle and right is for J1009 field with the $z=0.2279$
  galaxy. For the J0943 $z=0.5485$ galaxy, the {\OVI} has a large
  velocity spread that extends opposite to the rotational direction of
  the galaxy. The J1009 $z=0.2279$ galaxy is counter-rotating with
  respect to the bulk of the {\OVI} absorption.}
\label{fig:B2}
\end{center}
\end{figure*}
\begin{figure*}
\begin{center}
\includegraphics[scale=0.47]{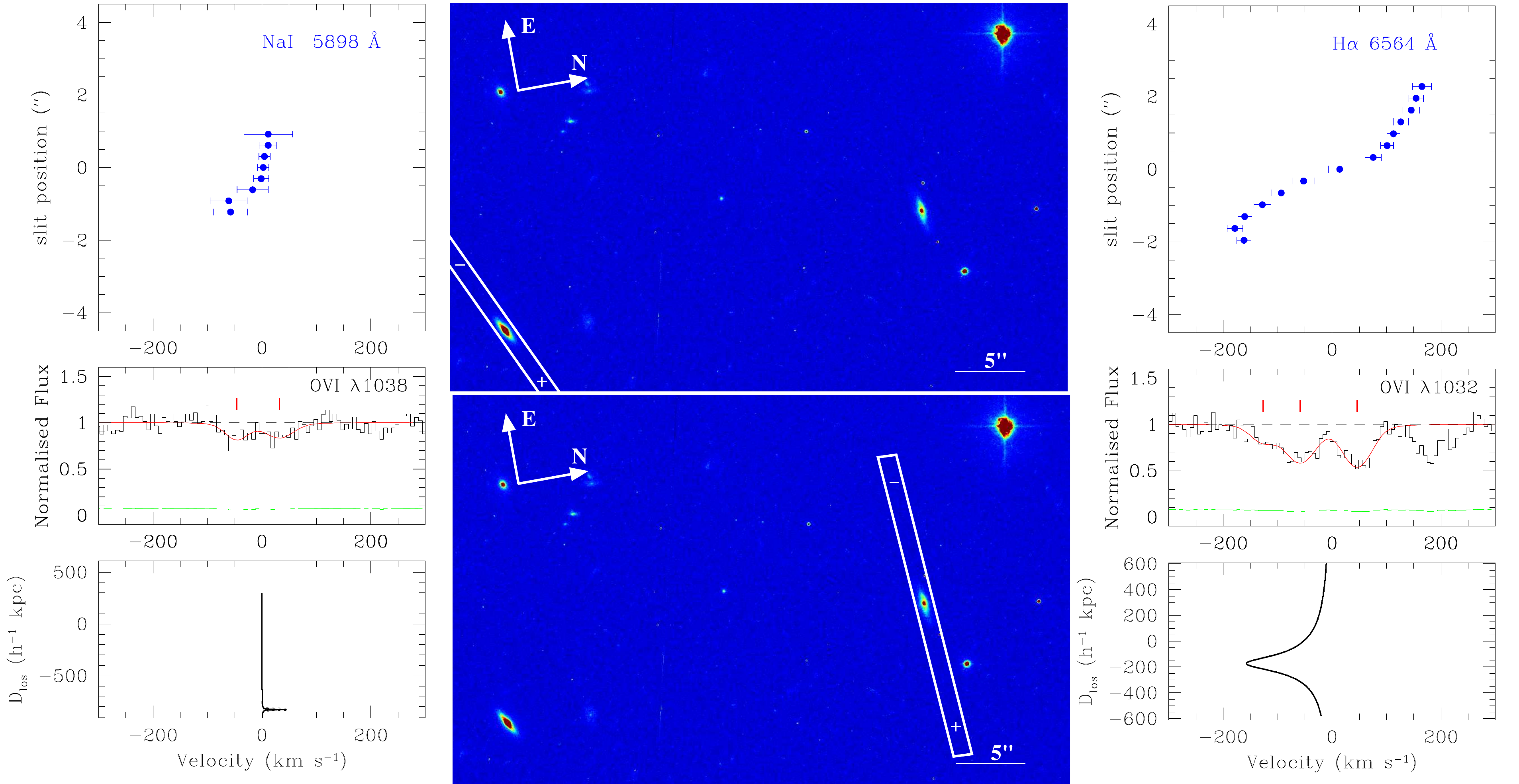}
\caption[angle=0]{Same as Figure~\ref{fig:A1} except that a 55$''
  \times $35$''$ {\it HST} in middle top and bottom panels is for the
  J1136 field.  The left panel is shown for the $z=0.2123$ galaxy and
  the right is shown $z=0.3193$ galaxy.  The {\OVI} for both galaxies
  spans boths sides of the systemic velocity, with little-to-no
  absorption at the systemic velocity. In both cases, the galaxies are
  counter-rotating with respect to roughly half of the {\OVI}
  absorption.}
\label{fig:B3}
\end{center}
\end{figure*}
\begin{figure*}
\begin{center}
\includegraphics[scale=0.47]{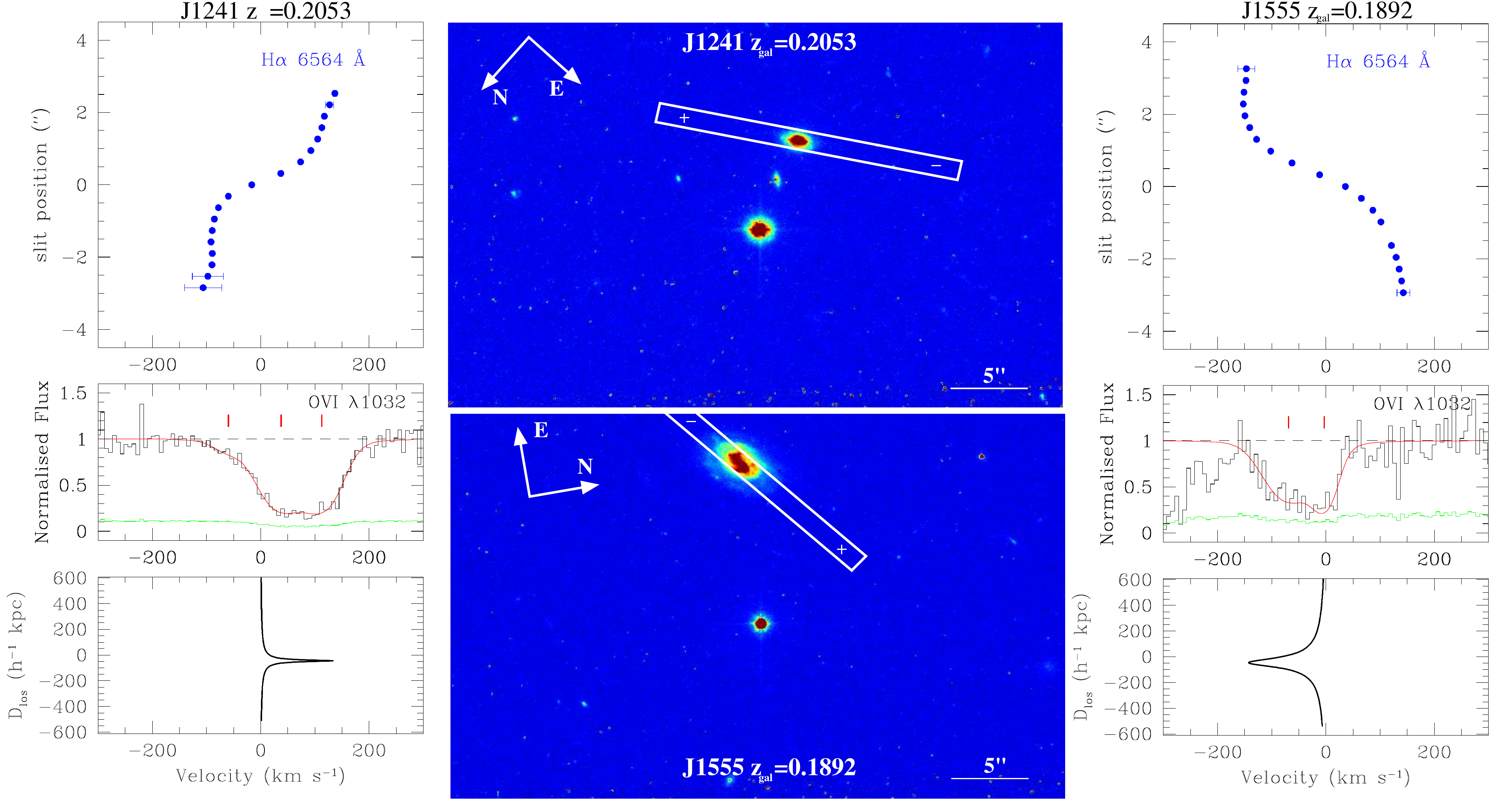}
\caption[angle=0]{Same as Figure~\ref{fig:A1} except for top-middle
  and left is for the J1241 field with the $z=0.2053$ galaxy and
  bottom-middle and right is for J1555 field with the $z=0.1892$
  galaxy.  In both cases, the galaxies are co-rotating with respect to
  the {\OVI} absorption.}
\label{fig:B4}
\end{center}
\end{figure*}
\begin{figure*}
\begin{center}
\includegraphics[scale=0.47]{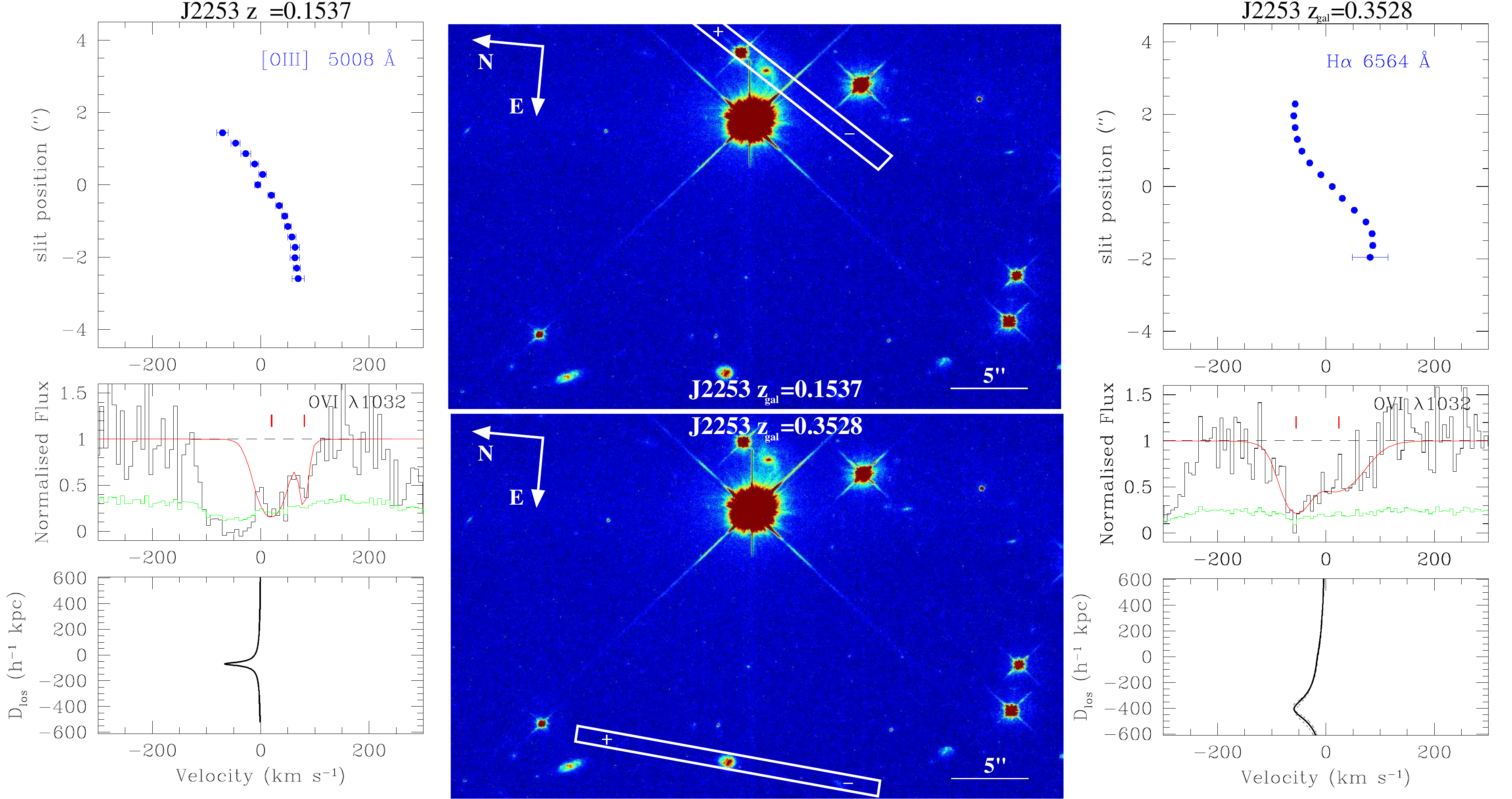}
\caption[angle=0]{Same as Figure~\ref{fig:A1} except the middle top
  and bottom panels is for the J2253 field.  The left panel is shown for
  the $z=0.1537$ galaxy and the right is shown $z=0.3528$ galaxy.  The
  {\OVI} for both galaxies spans boths sides of the systemic velocity,
  the galaxies are co-rotating with the highest optical depth {\OVI}
  absorption.}
\label{fig:B5}
\end{center}
\end{figure*}
\end{appendix}

\end{document}